\documentclass[twocolumn,onecolappendix]{aastex63}

\usepackage{makecell}
\usepackage{xcolor}
\definecolor{dgreen}{RGB}{26,148,49}
\definecolor{xlinkcolor}{cmyk}{1,1,0,0}
\hypersetup{linkcolor=xlinkcolor,citecolor=xlinkcolor,urlcolor=xlinkcolor}

\usepackage{amsmath,amsthm,amssymb,amsfonts}
\usepackage{graphics}
\usepackage{listings}
\usepackage{enumitem}
\usepackage{url}
\usepackage{appendix}
\usepackage{hyperref}
\usepackage{mathtools}

\usepackage[super]{nth}

\usepackage{float}
\usepackage{booktabs}

\newcommand{\N}{\mathbb{N}}
\newcommand{\Z}{\mathbb{Z}}

\begin{document}

\correspondingauthor{Kendrick Nguyen}
\email{knguyen7224@berkeley.edu}

\author[0009-0004-6209-6762]{Kendrick Nguyen}
\affiliation{Berkeley Center for Cosmological Physics, University of California, Berkeley, CA 94720,
USA}

\author[0000-0002-9330-8738]{Richard M. Feder}
\affiliation{Berkeley Center for Cosmological Physics, University of California, Berkeley, CA 94720,
USA}
\affiliation{Lawrence Berkeley National Laboratory, Berkeley, California 94720, USA}

\author[0000-0002-6503-5218]{Sean \surname{Bruton}}
\affiliation{California Institute of Technology, Pasadena, CA 91125, USA}

\author[0000-0001-5382-6138]{Uroš Seljak}
\affiliation{Berkeley Center for Cosmological Physics, University of California, Berkeley, CA 94720,
USA}
\affiliation{Lawrence Berkeley National Laboratory, Berkeley, California 94720, USA}

\title{A Multimodal Approach to Star--Galaxy Separation using SPHEREx Spectrophotometry and DESI Legacy Survey Imaging}
\begin{abstract}
Stellar contamination is a critical systematic for increasingly precise large-scale structure analyses from ongoing and next-generation surveys. Experiments targeting constraints on local primordial non-Gaussianity with $\sigma(f_{\rm NL}^{\rm loc}) \sim \mathcal{O}(1)$ demand sub-percent stellar contamination rates to avoid misidentifying spurious large-scale power induced by Galactic structure as true cosmological signal. In this work, we explore the use of multimodal models for star--galaxy separation, harnessing the information from both optical broad-band imaging data and SPHEREx near-infrared low-resolution spectrophotometry. The two modalities are integrated using contrastive learning, which projects image- and spectrum-based embeddings into a shared latent space. We find that classifiers trained on these transformed representations outperform those trained on the original embeddings and show less performance degradation when simpler classifiers are used. These results suggest that multimodal alignment organizes the embedding space along dimensions that are better suited to source classification. The improvement is particularly strong for image-based classification, which we connect to increased predictability of highly-discriminative infrared spectral features from the transformed image embeddings. Applying redshift error-based selections and extrapolating to the full SPHEREx footprint, we demonstrate that stellar contamination can be controlled at the sub-percent level across most of the extragalactic sky, with completeness tradeoffs largely confined to low redshift. Our work highlights the utility of multimodal methods for modern galaxy surveys such as SPHEREx and \emph{Rubin} LSST.  

\end{abstract}


\section{Introduction}
\label{sec:introduction}

Large-scale structure (LSS) surveys have emerged as one of the primary means of constraining fundamental cosmological parameters, with ongoing and next-generation experiments designed to measure or constrain the galaxy power spectrum, baryon acoustic oscillations, and primordial non-Gaussianity. Photometric surveys such as the Dark Energy Survey \citep[DES;][]{des}, the Kilo-Degree Survey \citep[KiDS;][]{kids}, and the forthcoming \emph{Rubin} Legacy Survey of Space and Time \citep[LSST;][]{rubin} are delivering imaging over thousands of square degrees, enabling characterization of samples with hundreds of millions of galaxies. Spectroscopic surveys, including the Dark Energy Spectroscopic Instrument \citep{desi}, complement these surveys with precise redshift measurements. As the precision of clustering measurements increases, the scientific return depends increasingly on the construction of well-characterized galaxy samples. Because contamination by stars and other non-cosmological sources introduces large-scale power that can obscure or mimic genuine cosmological signals, accurate star--galaxy separation is a foundational requirement for the target selection and sample construction of Stage-IV cosmology surveys. 

The Spectrophotometer for the History of the Universe, Epoch of Reionization, and Ices Explorer \citep[SPHEREx;][]{bock25} began survey operations in May 2025 and will assemble a sample of $\sim 800$M galaxies with redshift precision $\sigma_{z/(1+z)}<0.2$ \citep{feder23}. While SPHEREx aims to constrain local primordial non-Gaussianity with $\sigma(f_{\rm NL}^{\rm loc}) \sim \mathcal{O}(1)$\footnote{Assuming universality, i.e., $b_{\phi}=\delta_c b_1(1-p)$}, systematic variations from stellar contamination, which trace large-scale Galactic structure, can easily mimic or exceed primordial signals of this magnitude \citep{wen_sfb_systematics}. Accurately characterizing the effects of stellar contamination on SPHEREx $f_{\rm NL}^{\rm loc}$ constraints requires assessing galaxy purity for samples organized by redshift, redshift error, and potentially other observables. In general, the power spectrum measurements demand sub-percent stellar contamination rates across samples in the absence of additional mitigation \citep{weaverdyck2020, wen_sfb_systematics}.

Star--galaxy separation is commonly performed using optical imaging data, such as from the DESI Legacy Imaging Surveys (denoted DESI-LS throughout), where sources are distinguished through morphological modeling and broad-band colors \citep{LS, zhoulrg}. This paradigm is common across modern and next-generation imaging surveys such as the Vera C. Rubin Observatory LSST \citep{rubin}. Such methods are complemented by cross-matching to astrometric catalogs from \emph{Gaia} \citep{gaia}, which robustly identify stars by detecting proper motion and computing distances with parallax information. However, the effectiveness of these approaches is limited at fainter magnitudes, where astrometric measurements become incomplete or unavailable. In spectrophotometric surveys such as J-PAS \citep{jpas}, classification is traditionally performed through template fitting, in which observed spectral energy distributions are compared to stellar and galaxy models. More recently, machine learning approaches have been applied to multi-band photometry and low-resolution spectral data, including tree-based methods such as random forests and eXtreme Gradient Boosting (XGBoost) \citep{xgboost}, which operate on fluxes, colors, and morphological features \citep{sdss_rf, miniJPAS_stargal}. 

An alternative is to jointly represent astronomical sources using multiple data modalities. This perspective is motivated by the development of foundation models, which are trained on large and diverse datasets to learn general-purpose representations that can be adapted to downstream tasks \citep{aion}. Such representations may capture higher-level structure in the data and, in some cases, disentangle informative features from noise \citep{denoising_multimodal}. Extending this idea, multimodal foundation models learn representations from multiple observational views of the same source---such as spectra, images, and time-series photometry---with the goal of capturing the underlying physical properties that give rise to each modality. AstroCLIP \citep{astroclip} and SpecCLIP \citep{specclip} are examples of this paradigm, using contrastive learning to align different astronomical data modalities within a shared embedding space.

AstroCLIP particularly highlights the utility of spectral information for improving imaging representations. Quantities such as stellar mass, metallicity, and stellar-population age are more accurately recovered from image embeddings after they are restructured to emphasize information they share with the spectra. These results suggest that paired spectroscopy can reorganize image representations so that physically meaningful information already present in the imaging becomes more readily accessible for downstream models.

In this work, we explore the use of multimodal models to assess whether a shared embedding between the DESI Legacy Survey imaging \citep[DESI-LS;][]{LS} and SPHEREx spectrophotometry can improve the quality of star–galaxy separation. In doing so, we assess the information content of the SPHEREx data and its complementarity with broad-band imaging. More broadly, we seek to understand how and why alignment improves classification, both in terms of absolute performance and robustness.

Our work is organized as follows. In \S \ref{sec:datasets} we provide an overview of the multimodal contrastive learning framework we use and describe the SPHEREx and DESI-LS datasets, along with our cross-matched catalog. We then outline the  classification pipeline in \S \ref{sec:methods} and present our results in \S \ref{sec:results}. In \S \ref{sec:stellar_contam} we consider stellar contamination more specifically for the SPHEREx all-sky galaxy survey and estimate stellar contamination rates that incorporate redshift error-based selections. Lastly, we conclude and discuss future avenues for improvement in \S \ref{sec:conclusion}. 
\section{Alignment Framework and Datasets}
\label{sec:datasets}

\begin{figure*}
    \centering
    \includegraphics[width=0.52\linewidth]{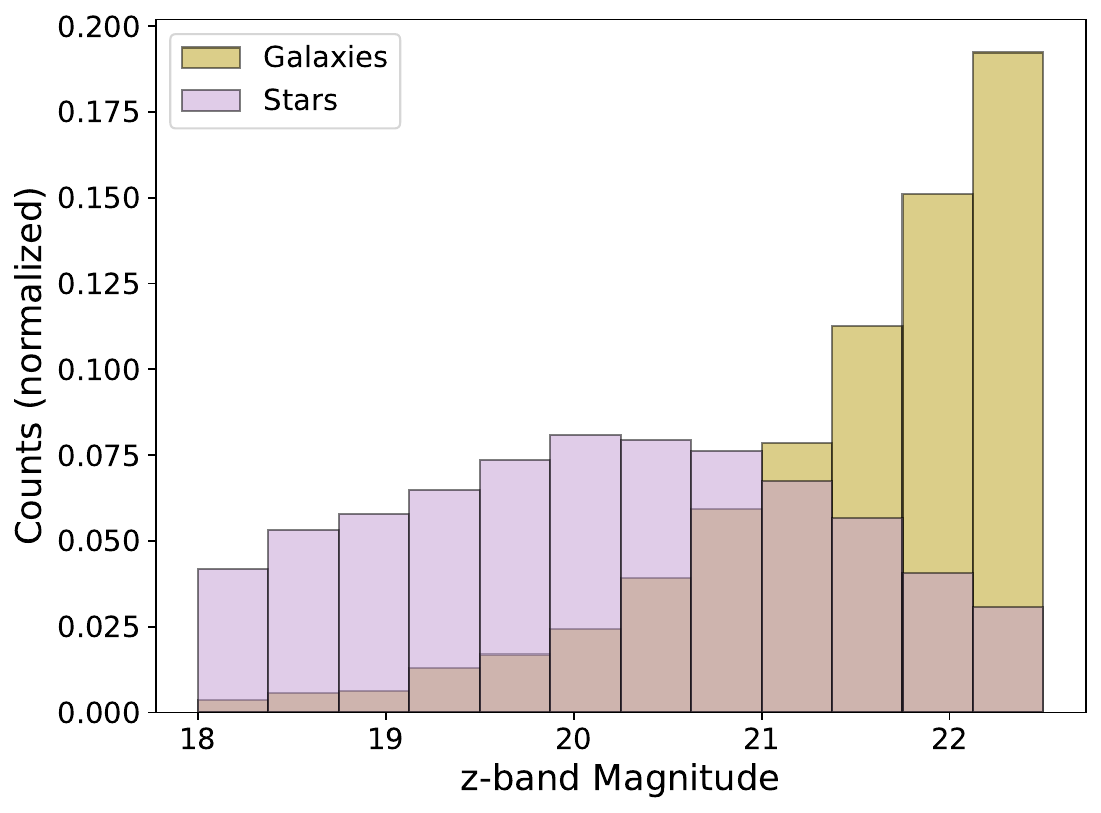}\includegraphics[width=0.46\textwidth]{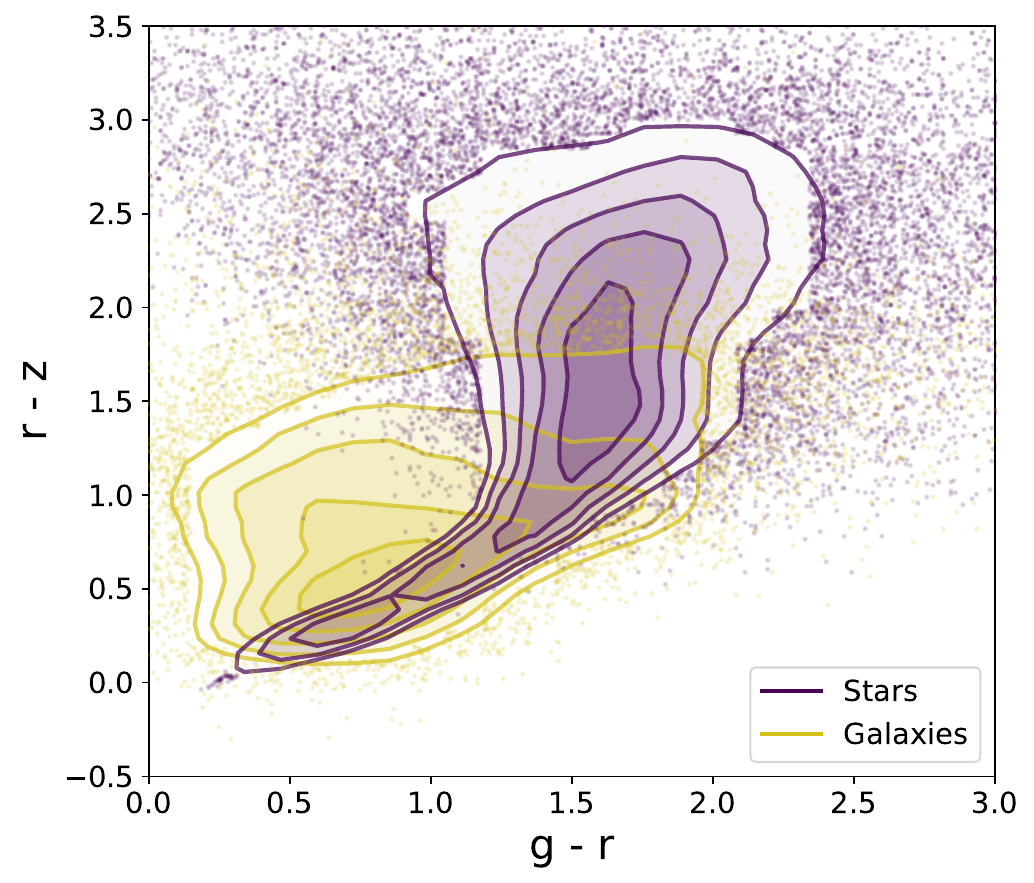}
    \caption{\textbf{Left}: Distribution of object counts for stars and galaxies as a function of z-band magnitude. \textbf{Right}: Color-color diagram of galaxies and stars. Contours are spaced at logarithmic intervals in source density. Stars and galaxies exhibit substantial overlap in optical broadband color-color space, with approximately $65\%$ of galaxies and $69\%$ of stars falling within the other's $95\%$ density region.}
    \label{fig:data_distribution}
\end{figure*}

\subsection{Overview of alignment framework}
\label{sec:alignment overview}
Our approach follows from the Contrastive Language–Image Pre-training (CLIP) algorithm \citep{clip}. CLIP aligns encoders from separate data modalities within a shared embedding space using a contrastive training objective. The objective function encourages embeddings of the same object across modalities to be drawn together in the shared embedding space, while embeddings of different objects are pushed apart. As a result, the learned representations emphasize information that is consistent across modalities while reducing sensitivity to modality-specific features.

The alignment objective is to maximize the mutual information between representations of the same object observed in two different modalities. In practice, however, mutual information is difficult to estimate directly for finite datasets \citep{info_limit}. CLIP instead minimizes the Information Noise Contrastive Estimation (InfoNCE; \citealt{infonce}) loss, which provides a variational lower bound on the mutual information, given by  
\begin{equation}
    \label{equation:info_nce}
    \mathcal{L}_{\text{InfoNCE}}(\mathbf{X}, \mathbf{Y}) = 
    -\frac{1}{K} \sum_{i=1}^K 
    \log \frac{
        \exp\!\left( S_C(\mathbf{x}_i, \mathbf{y}_i)/\tau \right)
    }{
        \sum_{j=1}^K \exp\!\left( S_C(\mathbf{x}_i, \mathbf{y}_j)/\tau \right)
    },
\end{equation}
where $\mathbf{x}_i$ and $\mathbf{y}_i$ denote the representations of the same object in modalities $\mathbf{X}$ and $\mathbf{Y}$, respectively, and $\tau$ is a temperature hyperparameter. The denominator sums over all samples $j$ within the minibatch of size $K$, comparing object $i$ to both its matched pair and non-matching samples in the opposite modality. The similarity metric $S_C$ quantifies the similarity between embeddings. CLIP uses the cosine similarity score given by  
\begin{equation}
    \label{equation:cosine_similarity}
    S_C(\mathbf{x}_i, \mathbf{y}_j) =
    \frac{\mathbf{x}_i^{\top} \mathbf{y}_j}{
    \lVert \mathbf{x}_i \rVert_2 \,
    \lVert \mathbf{y}_j \rVert_2
    }.
\end{equation}

In general, the representations of the two modalities do not share the same dimensionality. They must first each be projected into the shared embedding space before the alignment objective is computed. We detail this process of projecting our data representations into a shared embedding space in \S \ref{sec:align}.

\subsection{Training data}
This work makes use of the DESI-LS multi-band imaging and synthetic SPHEREx spectra. Because contrastive learning requires paired samples, each source in the imaging data must have an associated spectrum. In this section, we describe the process of obtaining these image-spectrum pairs. 

To cross-match and analyze the two datasets, we use infrastructure developed by the Multi-Modal Universe (MMU) project \citep{mmu}. The MMU dataset consists of 100 TB of data from astronomical sources in a variety of modalities, including multi-channel and hyper-spectral images, spectra, and multivariate time series. MMU additionally provides the code used to extract each dataset from its parent survey.

We assume a fiducial sample pre-selection using DESI Legacy Survey $z$-band magnitudes, retaining all sources with  $z_{\rm AB}<22.5$. This is slightly different than the magnitude-color selection used in \cite{huai25}. Figure \ref{fig:data_distribution} compares the distributions of stars and galaxies in $z$-band magnitude and color--color space.

\subsubsection{Synthetic SPHEREx galaxy spectra}
\label{sec:sphx_spectra}
We use synthetic SPHEREx spectra described in
\cite{feder23} and publicly available on Zenodo\footnote{\url{https://zenodo.org/records/11406518}}. In brief, the synthetic galaxy SEDs are derived from template fits to multi-band photometry of galaxies in the COSMOS 2020 catalog \citep{cosmos2020}, after which emission lines following semi-empirical relations with redshift, stellar mass, and galaxy type are added. We then convolve these SEDs with a set of 102 fiducial SPHEREx bandpass filters and add noise commensurate with the SPHEREx full-sky sensitivity. After selecting galaxies with $z_{\rm AB}<22.5$, our initial galaxy sample contains $31,\!149$ sources over the 1.27 deg$^2$ COSMOS footprint, or $\overline{n}_{\rm gal}=24500$ deg$^{-2}$. As the simulated SPHEREx spectra are based on real sources in the COSMOS field, we can cross-match to DESI-LS images of the same sources. 

\subsubsection{DESI Legacy Survey Imaging}
\label{sec:desils_imaging}
The DESI Legacy Imaging Surveys \citep{LS} provide wide-field optical imaging in $g$, $r$, and $z$ bands over the extragalactic sky. DESI-LS comprises data from the DECam Legacy Survey (DECaLS), which covers mostly the southern sky, and the Beijing–Arizona Sky Survey (BASS) and Mayall z-band Legacy Survey (MzLS) for the North. All imaging is processed through a uniform pipeline and modeled with \texttt{The Tractor} \citep{tractor}, producing consistent photometry, morphological classifications, and co-added images. Specifically, we use imaging from DR10 of DESI-LS. The COSMOS field is fully covered by the DECaLS footprint, and all image cutouts used in this work are sourced from this survey. The co-added images have a pixel side length of 0.262\arcsec, while the mean PSF full widths half maximum (FWHM) are 1.29\arcsec, 1.18\arcsec, and 1.11\arcsec\ for $g$-, $r$-, and $z$-bands, respectively.

We use a modified version of the image cutout extraction pipeline from MMU\footnote{\url{https://github.com/MultimodalUniverse/MultimodalUniverse/tree/main/scripts/legacysurvey}} to obtain galaxy images corresponding to our simulated spectra. We generate custom sweep files containing the positions of sources in our SPHEREx galaxy catalog and relax the magnitude cut from $z_{\rm AB} < 21$ to $z_{\rm AB} < 22.5$, while retaining the same quality cut flags for bad pixels. These sweep files are then passed to the cutout extraction script to produce 144 $\times$ 144 pixel postage stamps, which are subsequently cross-matched with the SPHEREx spectra. We exclude sources whose DESI-LS cutouts contain bad pixels using the same quality-mask criteria as in MMU \citep{mmu}, which removes 24.65\% of our initial galaxy sample.


\subsubsection{Synthetic stars}
\label{sec:synthetic_stars}
As we lack a well-characterized catalog of real stars with cross-matched imaging and public SPHEREx spectra, we use simulated stars and image injections to generate our synthetic dataset. The mocks use \texttt{pystellib}\footnote{\url{https://github.com/mfouesneau/pystellibs}}, which computes synthetic stellar spectra based on properties such as $T_{\rm eff}$, surface gravity $\log g$, and metallicity approximated by [Fe/H]. The stellar parameters for our simulated sample are drawn from the \texttt{BaSeL v2.2} library \citep{basel}. We derive the stellar density by populating HEALPix \texttt{NSIDE}=32 cells using the spatial distribution from \texttt{Galaxia}, which uses a stellar disk + halo model \citep{galaxia}. Because the synthetic SEDs describe the intrinsic stellar spectra, we redden them using a uniform $E(B-V) = 0.02$ with \citet{Fitzpatrick1999} Milky Way dust law, matching the measured extinction in the COSMOS field \citet{Scoville2007COSMOS}:
\begin{equation}
    m_{\mathrm{obs}} = m_{\mathrm{sim}} + R_{\lambda}\,E(B - V).
\end{equation}
We use the same extinction coefficients as the DESI Legacy Imaging Survey DR10 for the $g$, $r$, and $z$ bands, namely $R_g = 3.214$, $R_r = 2.165$, and $R_z = 1.211$ \citep{LS,LS_extinction}. The synthetic stars we use for training and testing cover a single 3.3 deg$^2$ HEALPix tile centered on the COSMOS field. After applying our $z_{\rm AB}<22.5$ magnitude cut, we are left with $62,\!558$ stars, or $\overline{n}_{\rm star}=18,\!956$ deg$^{-2}$.


As the synthetic stellar spectra do not correspond to real sources on the sky, we inject stars into the DESI-LS imaging as point-like sources, using empirical point spread function (PSF) models derived from the DESI-LS imaging pipeline. These models are generated with \texttt{PSFEx} and capture the spatially varying, non-Gaussian structure of the instrument response \citep{LS,psfex}. We convolve the simulated SEDs with the DESI-LS broadband filter transmission curves to synthesize $g$-, $r$-, and $z$-band fluxes for each object. We then assign ``blank-sky" injection positions within the COSMOS field located at least $1.5$\arcsec\ from any source in the pre-magnitude-cut simulated galaxy catalog and from any other injected star. This pre-magnitude-cut catalog contains 54,013 sources with $i<25$, ensuring that injected stars do not overlap with fainter galaxies outside the final $z_{\rm AB}<22.5$ sample.

\section{Methods}
\label{sec:methods}
We implement a framework based on CLIP through the following procedure:
\begin{enumerate}
    \item Train a spectrum encoder,
    \item Align the spectrum encoder with a pre-trained image encoder using the CLIP method,
    \item Use aligned embeddings of objects with known labels to train an XGBoost classifier.
\end{enumerate}
The following subsections describe each component of this procedure in detail.
\subsection{AstroCLIP and Image Encoder}
\label{sec:image_encoder}
AstroCLIP \citep{astroclip} is designed to jointly embed DESI-LS images and DESI spectroscopy of galaxies. Two transformer-based encoders are pretrained on the separate modalities. The embeddings from both encoders are then projected into a shared latent space and aligned by minimizing the InfoNCE loss (Eq. \eqref{equation:info_nce}), as in \citet{clip}.

\citet{astroclip} show that such multimodal alignment can produce representations that perform competitively on a range of downstream tasks, including photometric redshift estimation and galaxy property inference, demonstrating the promise of contrastive multimodal learning in astronomical applications.

We deploy the pre-trained unaligned image encoder from AstroCLIP (AstroDINO) in the alignment step with our spectrum encoder. Details of the AstroDINO architecture and pre-training are provided in the original work \citep{astroclip}. To overview, inputs of $144 \times 144$ pixel images with 3 channels $(g,r,z)$ are divided into 144 total $12 \times 12 \times 3$ non-overlapping patches. Properties within each patch are encoded in vectors of length 1024, from which a 1024-dimensional global summary vector is extracted. The alignment dataset used in our work differs from the AstroDINO pre-training set in two important ways: we include both stars and galaxies rather than galaxies alone, and we use a broader magnitude range rather than restricting the sample to sources with $20 < z_{\rm AB} < 21$.

\subsection{Spectrum Encoder}
\label{sec:spectrum_encoder}

\begin{figure*}
    \centering
    \includegraphics[width=\textwidth]{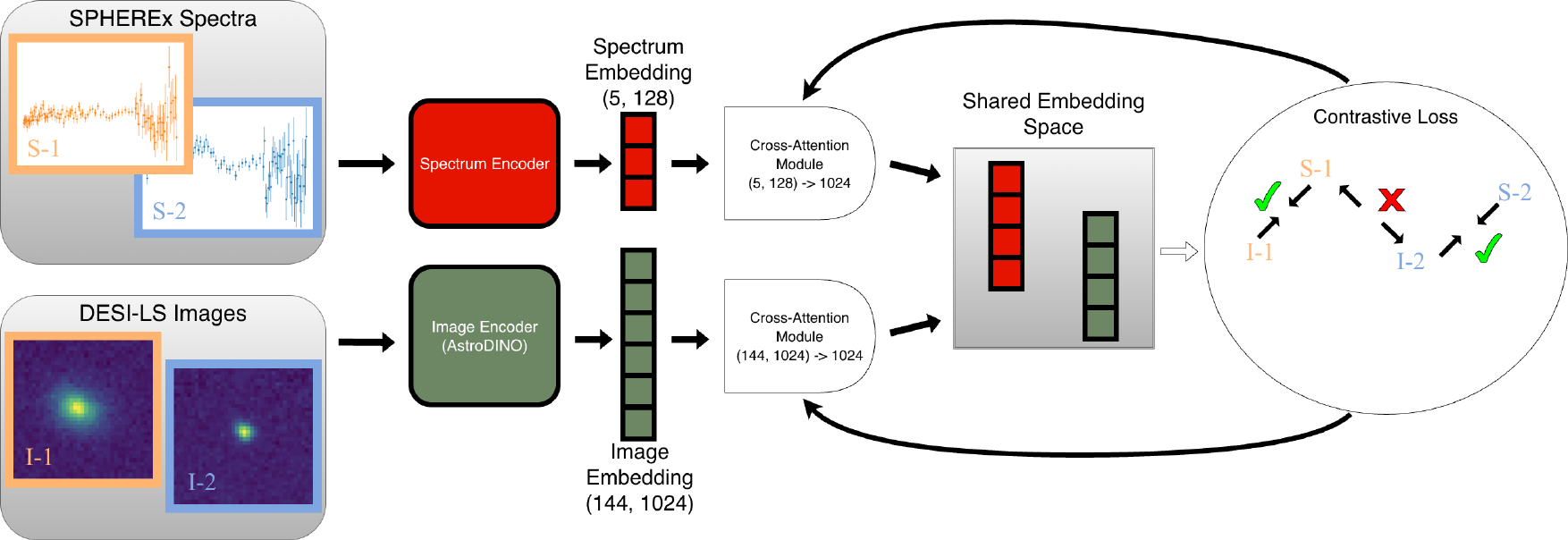}
    \caption{Schematic of alignment framework. The encoders output representations of their respective data, which are then projected into a shared embedding space. A contrastive loss encourages the projection heads to push image-spectrum pairs of the same object together and mismatched pairs apart.}
    \label{fig:schematic}
\end{figure*}

We use a 1D convolutional encoder consisting of four sequential convolutional layers with 1, 32, 64, and 128 filters, respectively, allowing progressively deeper layers to capture increasingly complex patterns in the spectral data. Each convolutional layer is followed by max-pooling with a kernel size of 5, which aggregates information across neighboring wavelength bins while progressively downsampling the spectrum. The resulting feature representation is then passed through a small multi-layer perceptron (MLP) that compresses it into a low-dimensional latent vector. The spectrum encoder forms the first half of a standard autoencoder, mapping the 102-channel photometry to a 6-dimensional latent space. This low-dimensional latent representation acts as an information bottleneck, encouraging the encoder to compress the input into a compact set of features that capture the most salient structure of the spectrum while discarding redundant information and noise. A decoder forms the second half of the autoencoder, and reconstructs the original spectrum from this latent representation. 

We train the autoencoder on synthetic deep-field spectra, where the higher signal-to-noise ratio (SNR) allows the encoder to learn intrinsic spectral structure with minimal contamination from observational noise. Relative to the fiducial survey depth, the deep-field spectra have flux uncertainties lower by a factor of $\sim 7.07$, corresponding to approximately $50\times$ lower noise variance. Each deep-field spectrum has a corresponding counterpart in the fiducial-depth set, where both are generated from the same underlying noiseless spectrum but with different levels of observational noise added. In addition, the encoder is trained solely on galaxy spectra, rather than on a mixed sample of stars and galaxies. Training and validation are done using the full magnitude-cut sample with an $80/20$ split.

To train the autoencoder, we minimize an inverse-variance-weighted $\chi^2$ reconstruction loss,
\begin{equation}
L_{\rm recon}
=
\sum_{i=1}^{N} \frac{\left(x_i - \hat{x}_i\right)^2}{\sigma_i^2}
,
\end{equation}
where $x_i$ and $\hat{x}_i$ denote the input and reconstructed fluxes in spectral channel $i$, respectively, and $\sigma_i$ is the corresponding flux uncertainty. Prior to training, each spectrum is normalized by its mean flux across all channels.

At the end of training, the mean sample $\chi^2$ per channel on the deep-field validation set is $\simeq 1.18$. A potential concern is that the relatively small 6-dimensional bottleneck may restrict the amount of spectral information retained by the autoencoder. To assess this, we train an otherwise identical model with a 12-dimensional latent space, which achieves mean sample $\chi^2$ per channel $\simeq 1.10$, corresponding to a $\sim6.8\%$ reduction in reconstruction loss relative to the 6-dimensional model. This suggests that increasing the bottleneck dimensionality allows more information to be preserved in the final compressed representation.

However, the difference is not consequential for our application, as the downstream alignment with the image encoder does not rely on the final latent representation itself. Rather than using the latent vector produced by the autoencoder, we extract intermediate features after the input spectrum has passed through three convolutional layers and two max-pooling layers. These features have shape $(5,128)$, where each of the five vectors represents a learned embedding of a contiguous segment of the original spectrum. In this way, the spectrum is represented as a sequence of localized vectors rather than a single global vector. This representation is analogous to the set of vectors produced by the image encoder, from which its global summary vector is produced. We adopt this strategy for compatibility with the AstroCLIP alignment architecture. To test how the bottleneck dimensionality affects the information content within the intermediate features, we evaluated the downstream classifier, described in $\S\ref{sec:downstream_class}$, on intermediate features of deep-field and fiducial-depth spectra extracted from models with 6- and 12-dimensional bottlenecks. We find that the two models exhibit nearly identical star--galaxy classification performance, achieving AUCs of 0.9996 for both models on deep-field spectra, and 0.9666 and 0.9681 for the 6- and 12-dimensional models, respectively, when evaluated on fiducial-depth spectra.

For the fiducial survey depth spectra used in the downstream classification tasks, the autoencoder achieves median per-channel reconstruction losses of 0.99 for galaxies and 1.10 for stars, while the corresponding mean losses are 1.00 and 1.27. The larger discrepancy between the mean and median stellar loss indicates a subset of stars that reconstruct particularly poorly, consistent with stars being out-of-distribution relative to the training set.

\subsection{Alignment of image and spectrum embeddings}
\label{sec:align}

We align the image and spectrum encoders by projecting their respective representations into a shared embedding space, as illustrated schematically in Figure \ref{fig:schematic}. Following AstroCLIP, we perform these projections using multi-head cross-attention modules. Here, ``cross-attention" refers to how the module aggregates and reweights information across the set of input vectors within each modality, rather than between the image and spectrum branches.

For the image branch, a cross-attention module uses information across the set of 144 vectors output by the image encoder to generate a single vector. From this, an MLP with a residual connection produces the final 1024-dimensional embedding. We apply an analogous transformation to the intermediate features of the spectrum encoder. A separate cross-attention module and residual connection MLP use context aggregated from the 5 128-length vectors output by the spectrum encoder to similarly produce an embedding with 1024 dimensions. 

For this step, we freeze the weights of the encoders and only train the two cross-attention modules and residual MLP layers on the alignment objective. These projection heads are therefore responsible for identifying and reweighting information from their respective modalities in order to align the resulting embeddings.

We use a batch size of 1024 and train for 50 epochs using the AdamW optimizer, which decouples weight decay from the gradient update \citep{adamw}. Following \citet{astroclip}, we fix the temperature hyperparameter $\tau = 15.0$ from Eq. \eqref{equation:info_nce}. For these hyperparameters, the training loss decreases smoothly and reaches a stable plateau by the end of training.

We construct the training and validation sets by splitting the full magnitude-cut sample of star and galaxy image-spectrum pairs into an $80/20$ split, while preserving the class proportions in both sets.

\begin{figure*}[t]
    \centering
    \includegraphics[width=\textwidth]{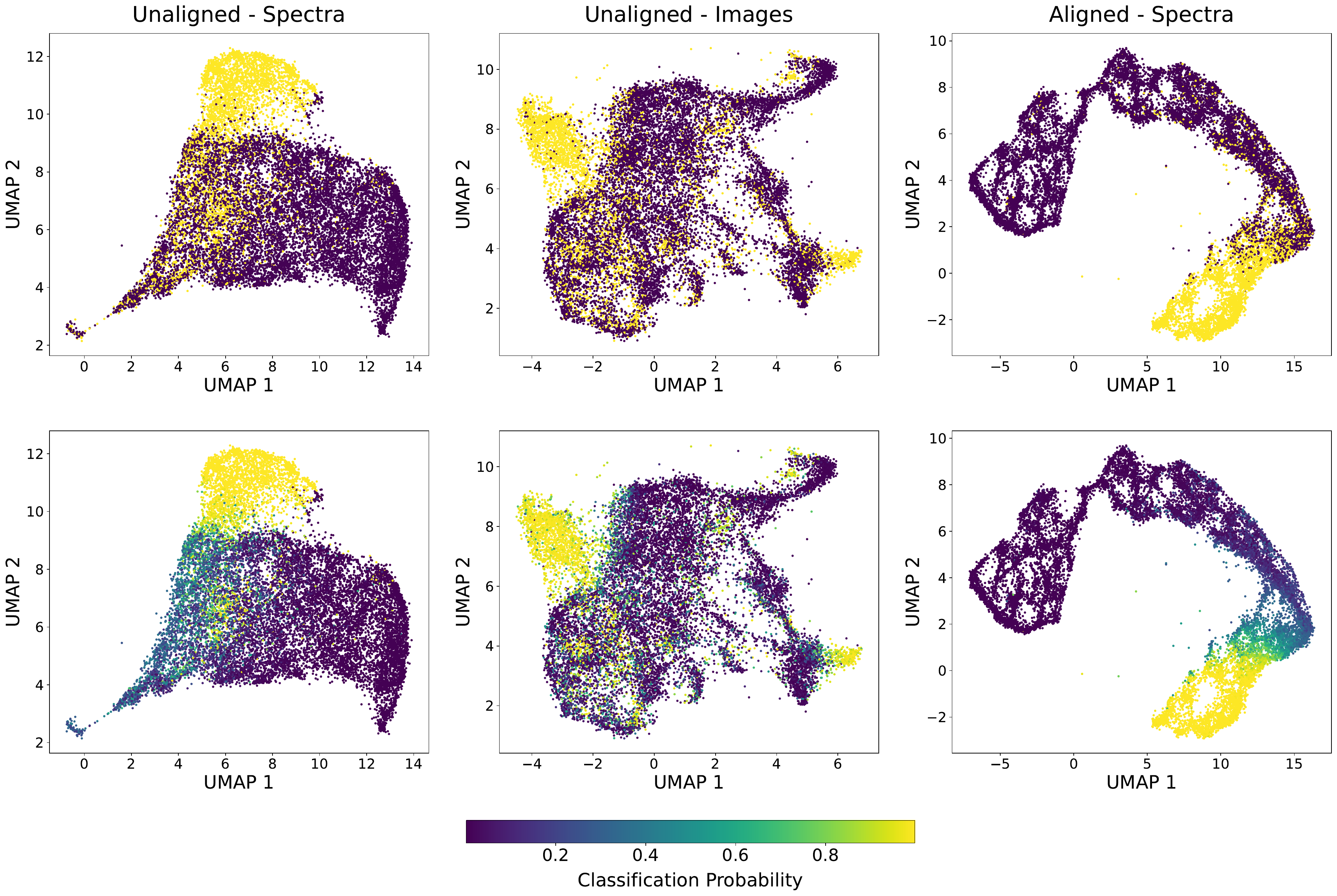}
    \caption{UMAP visualization of encoder embedding spaces. \textbf{Left:} unaligned spectrum encoder. \textbf{Middle:} unaligned image encoder. \textbf{Right:} spectrum encoder after contrastive alignment with the image encoder. 
    In the top row, point are colored by their ground-truth labels, with galaxies shown in yellow and stars in purple. In the top bottom, they are colored by the galaxy probability assigned by the XGBoost classifier. The aligned embedding exhibits a more coherent separation between low- and high-probability regions than either unaligned embedding. Embeddings of the aligned image encoder (not shown) are nearly identical to those of the aligned spectrum encoder; a comparison is provided in Appendix \ref{appendix:alignment visualization}.}
    \label{fig:class_proba_umap_comparison}
\end{figure*}

\subsection{Classification technique}
\label{sec:downstream_class}
We then train a classifier on the image and spectrum embeddings described in earlier subsections. Our focus in this work is to determine the impact of contrastive alignment on star--galaxy separation, and so we train separate models for the unaligned and aligned embeddings. We use XGBoost, a gradient-boosted decision tree algorithm that combines an ensemble of ``weak learners" (in our case, decision trees) to produce predictions for a given downstream task \citep{xgboost}. Unlike standalone decision trees or random forests, XGBoost builds trees sequentially, with each new tree fit to the residual errors of the previous ensemble. XGBoost uses both L1 and L2 regularization to yield robust predictions given the training data. 

For our default XGBoost model, we use an ensemble of 200 decision trees with a maximum depth of 6. Training the classifier itself is efficient, and we use early stopping based on convergence of the validation set loss. In \S \ref{sec:results}, we additionally evaluate a set of weaker classifiers to assess the sensitivity of the unaligned and aligned embeddings to classifier strength. Our XGBoost model returns star and galaxy class probabilities for each object, with the two probabilities summing to unity. The choice of $p_{\rm gal}^{\rm thresh}$ for a given analysis can be adjusted if higher purity or completeness is desired, which we explore in \S \ref{sec:forecast_contam_sky}.

\section{Results}
\label{sec:results}

In this section, we compare the classification performance of the aligned and unaligned models. In addition to the encoder-based approaches, we include a baseline in which XGBoost is trained directly on the raw spectra. In this section, we determine class labels using a classification probability threshold of $p_{\rm thresh} = 0.5$.

\subsection{Latent space visualization} 
We use Uniform Manifold Approximation and Projection \citep[UMAP; ][]{umap}, a dimensionality reduction technique, to visualize our high-dimensional data embeddings in a two-dimensional space. UMAP assumes that the data are distributed on a locally connected Riemannian manifold with approximately uniform density, and seeks to preserve both local and global topological structure under projection to a lower-dimensional space. Because the observable properties of stars and galaxies vary as smooth, continuous functions of underlying physical parameters (e.g., temperature, metallicity, redshift, and star formation rate), these assumptions are well motivated for the embeddings considered here.

In Figure \ref{fig:class_proba_umap_comparison} we compare the organization of the embeddings before and after alignment. The top row shows the true class labels projected into each embedding space, with galaxies in yellow and stars in purple. In the unaligned spectrum embeddings (left), galaxies and stars occupy distinct regions but have substantial overlap near the class boundary. The unaligned image embeddings (middle) show galaxies distributed across several scattered clumps interspersed with stars. In contrast, the aligned spectrum embeddings (right) exhibit a much clearer separation between the two classes, with overlap between galaxies and stars confined to a substantially smaller region of the latent space. In Appendix \ref{appendix:physical_properties}, we show that the aligned representation also retains clear variations with respect to galaxy redshift and physical properties, such as stellar mass and H$\alpha$ line equivalent width. 

The bottom row shows the galaxy probabilities assigned by our XGBoost classifiers, trained on the same three sets of embeddings. The unaligned spectrum embedding space shows a diffuse decision boundary with a large region of overlapping intermediate probabilities, while the unaligned image embeddings have comparatively weak class organization. In the aligned embedding space, the transition in classification probability is substantially smoother and more localized, indicating that alignment produces a more coherent representation in which the two classes are more readily separable.

We note here that the shared aligned space is visualized using embeddings produced by the aligned spectrum encoder. Because alignment encourages spectrum and image inputs of the same object to occupy nearby locations in the shared space, the aligned image embeddings have near-identical structure. However, some differences between the two persist, which we investigate further in Appendix~\ref{appendix:alignment visualization}.

While UMAP provides valuable qualitative insight into the organization of high-dimensional embeddings, it is not guaranteed to faithfully preserve global geometric relationships \citep{umap}. As a result, apparent separability in the two-dimensional projection may not directly correspond to quantitative classification performance. Likewise, classes that appear substantially overlapped in the UMAP projection may remain well separated in the original embedding space, where higher-dimensional structure is retained. We therefore turn to quantitative classification experiments in the following subsection to evaluate the true separability of the learned representations.
\subsection{Classification performance}
\subsubsection{Metrics}
We use purity and completeness to evaluate classification performance. We define galaxy purity as
\[
\mathrm{Purity}_{\mathrm{gal}} =
\frac{N_{\mathrm{TP}}}
{N_{\mathrm{TP}} + N_{\mathrm{FP}}},
\]
where \(N_{\mathrm{TP}}\) is the number of true galaxies correctly classified as galaxies, and \(N_{\mathrm{FP}}\) is the number of stars incorrectly classified as galaxies. We define galaxy completeness as
\[
\mathrm{Completeness}_{\mathrm{gal}} =
\frac{N_{\mathrm{TP}}}
{N_{\mathrm{TP}} + N_{\mathrm{FN}}},
\]
where \(N_{\mathrm{FN}}\) is the number of true galaxies incorrectly classified as stars. Stellar purity and completeness are defined analogously.

When computing class purity in this section, we re-weight the star and galaxy samples so that their relative abundances match those observed in the COSMOS field. This involves correcting for the different effective sky areas of the two samples ($A_{\rm star}=3.3$ deg$^2$ vs. $A_{\rm gal}=1.27$ deg$^2$), and the DESI-LS image-quality cuts described in \S \ref{sec:desils_imaging}. In addition, we apply a correction to account for stellar density mismatch between our simulations and observed data. By comparing mock stellar densities against \emph{Gaia} star counts for the same $G<20$ magnitude selection, we find that our mocks overpredict stellar counts by $\sim 50\%$ near the COSMOS field\footnote{This difference is driven in part by additional rejection of \emph{Gaia} sources near very bright stars. We caution that this correction is approximate and defer a full comparison of simulations and data to future work.}. In \S \ref{sec:forecast_contam_sky}, we see how the purity varies over the full extragalactic footprint and as a function of classification probability threshold $p_{\rm thresh}$. 


\begin{table*}
    \centering
    \begin{tabular}{lcccc}
    \toprule
    Data Representation & Stellar Purity & Stellar Completeness & Galaxy Purity & Galaxy Completeness \\
    \midrule
    \multicolumn{5}{c}{\textit{Spectrum-Based Classification}} \\
    \midrule
    Raw Spectra & 0.6818 & \textbf{0.9767} & \textbf{0.9840} & 0.7586 \\
    Unaligned Spectrum Embeddings & 0.6836 & 0.9554 & 0.9727 & 0.7626 \\
    Aligned Spectrum Embeddings & \textbf{0.6917} & 0.9733 & 0.9821 & \textbf{0.7712} \\
    \midrule
    \multicolumn{5}{c}{\textit{Image-Based Classification}} \\
    \midrule
    Unaligned Image Embeddings & 0.7559 & 0.9693 & 0.9810 & 0.8349 \\
    Aligned Image Embeddings & \textbf{0.9308} & \textbf{0.9831} & \textbf{0.9908} & \textbf{0.9614} \\
    \bottomrule
    \end{tabular}
    \caption{Classification results for the $z_{\rm AB}<22.5$ SPHEREx-like sample. Metrics are computed using a probability threshold of $p_{\rm gal}^{\rm thresh}=0.5$. The top section shows performance when classification is performed using spectroscopic information, while the bottom section shows performance using image-derived embeddings. These results do not include additional selections based on recovered photo-$z$ precision, which we detail in \S \ref{sec:stellar_contam}.}
    \label{tab:class_results}
\end{table*}
    
\subsubsection{Pre- and Post-Alignment Classification Results}
\begin{figure*}
    \centering
    \includegraphics[width=0.48\textwidth]{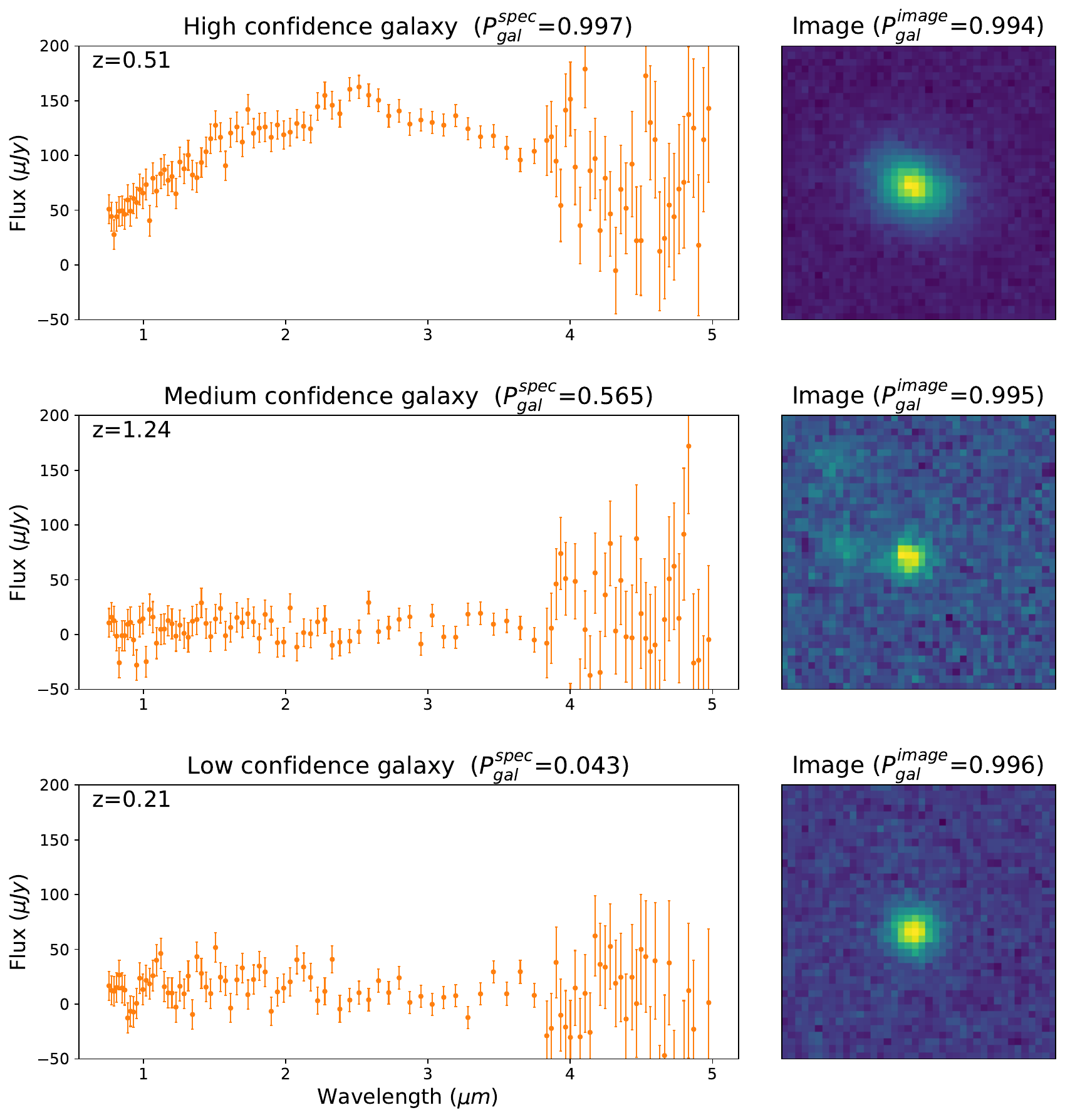}%
    \hspace{0.02\textwidth}
    \includegraphics[width=0.48\textwidth]{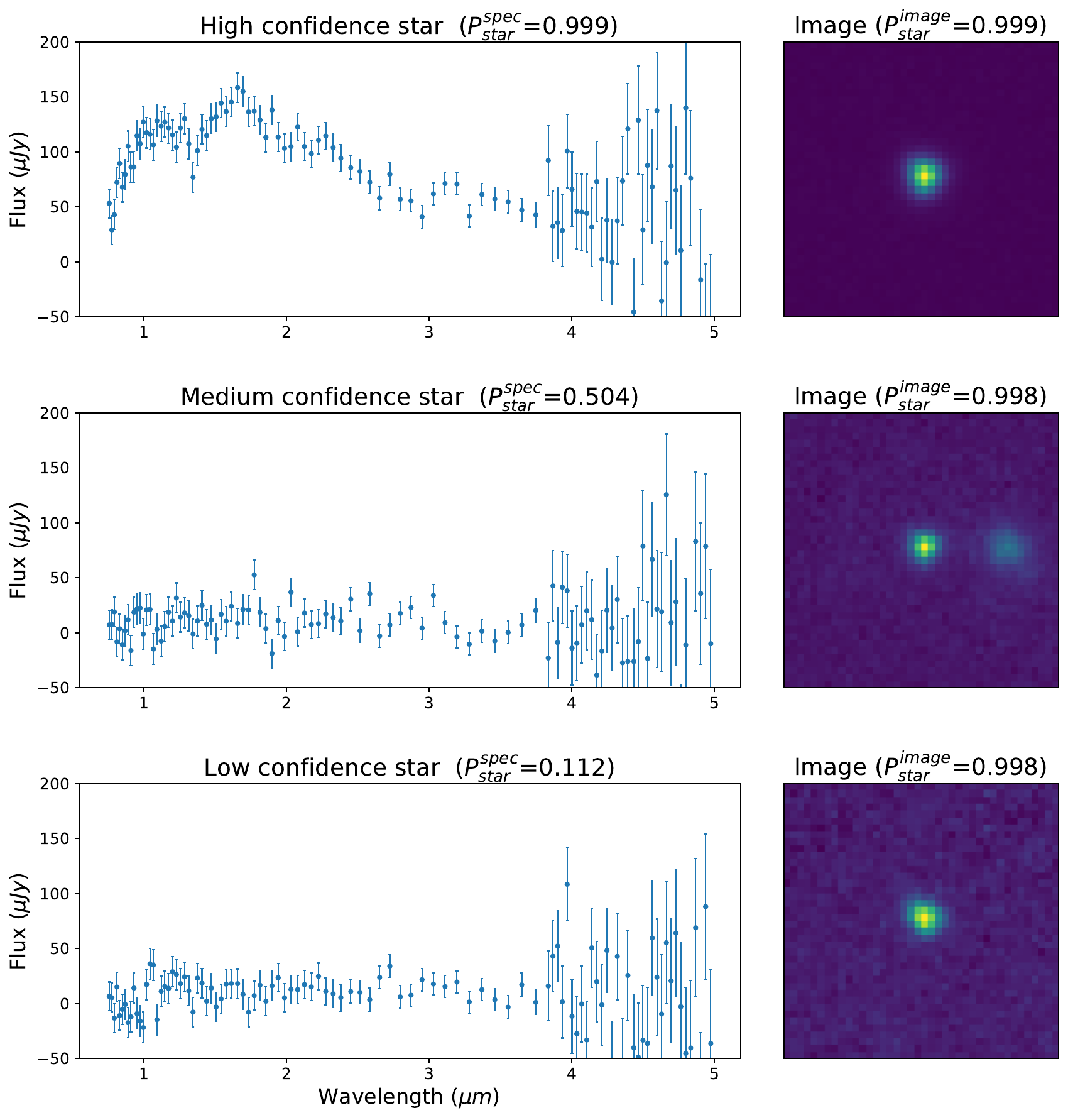}
    \caption{Examples of spectra with high, medium, and low class probabilities assigned by the XGBoost classifier trained on embeddings of the aligned spectrum encoder. The central $40 \times 40$ pixels of the corresponding $z$-band images are also shown, together with the class probabilities assigned by an XGBoost classifier trained on embeddings of the aligned image encoder. These examples illustrate how sources with uncertain spectral classifications can nevertheless remain visually distinctive, leading to higher-confidence predictions from the image-based classifier. They demonstrate how imaging information can, in some cases, help resolve ambiguities when spectral features are weak or difficult to distinguish. \textbf{Left:} Galaxies. \textbf{Right:} Stars.}
    \label{fig:spec_img_proba}
\end{figure*}
\begin{figure}
    \centering
    \includegraphics[width=0.48\textwidth]{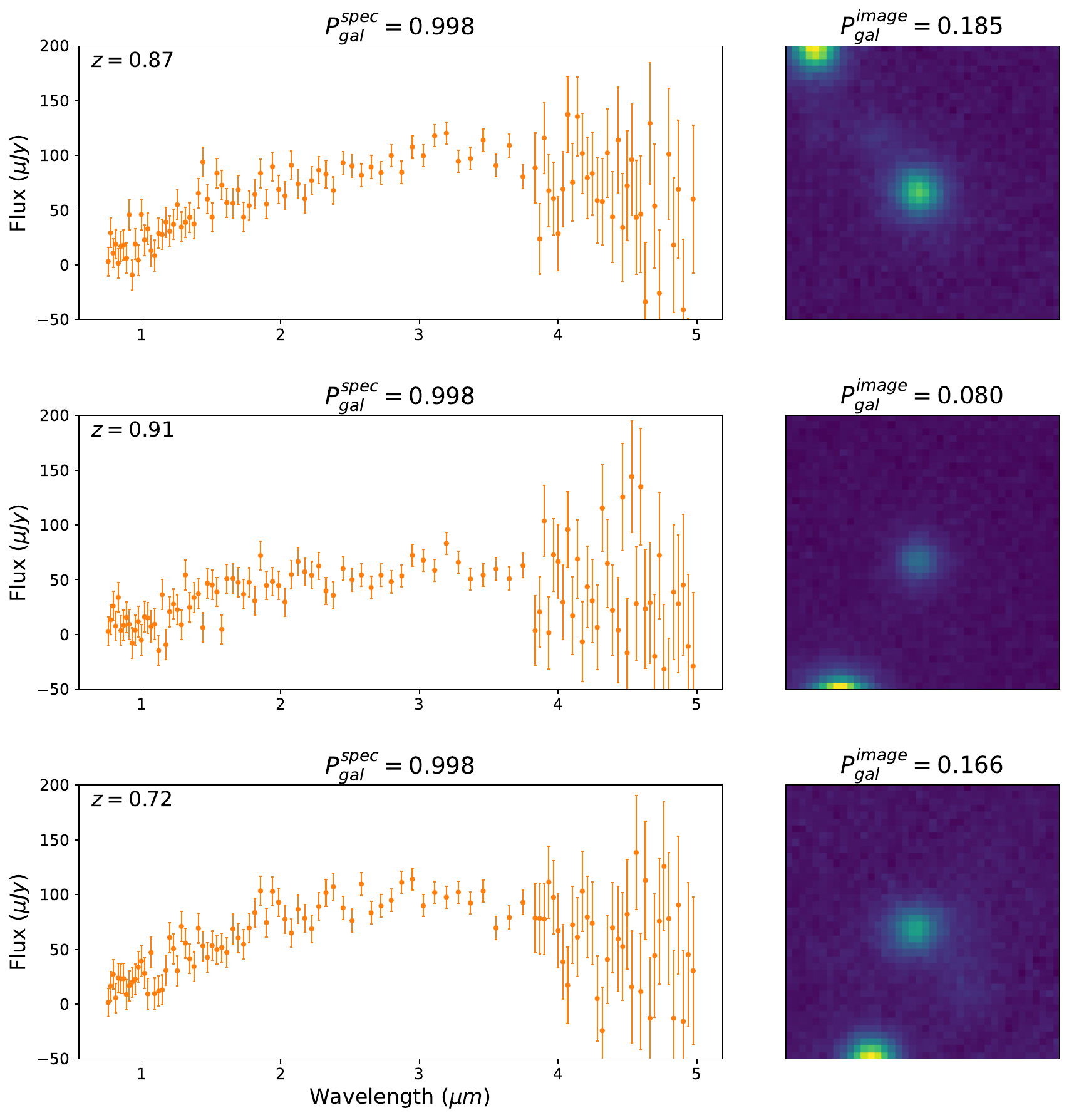}
    \caption{Examples of galaxies for which the spectrum-based classifier assigns a high galaxy probability while the image-based classifier assigns a low galaxy probability. Images show the central $40 \times 40$ pixels of the full cutout for easier visualization. Probabilities are produced by XGBoost classifiers trained on unaligned spectrum and image embeddings, respectively.}
    \label{fig:high_spec_low_img}
\end{figure}
\begin{figure*}
    \centering
    \includegraphics[width=\textwidth]{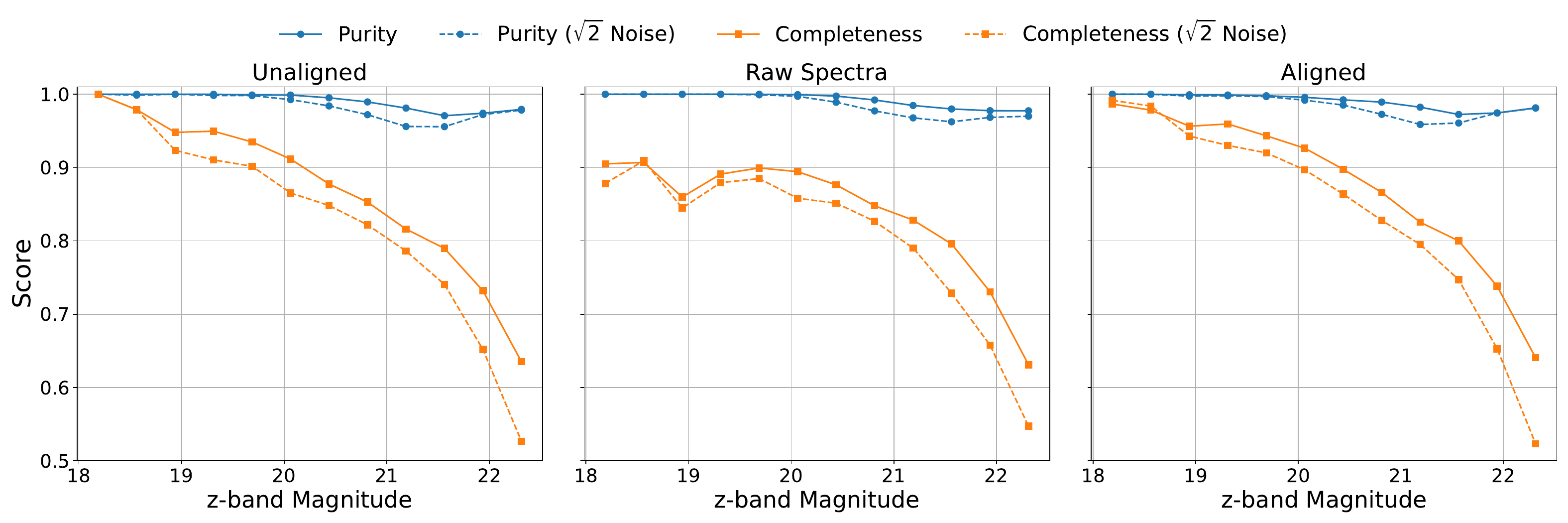}
    \caption{Galaxy purity and completeness of the XGBoost classifier using embeddings from the unaligned spectrum encoder (left), classification performed directly on mock SPHEREx spectra (middle), embeddings from the aligned spectrum encoder (right). We show results for classifiers evaluated on both the fiducial noise level test data (solid curves) and spectra with $\sqrt{2}\times$ higher flux uncertainties (dashed).}
    \label{fig:metric_comparison}
\end{figure*}
\begin{figure}
    \centering
    \includegraphics[width=\linewidth]{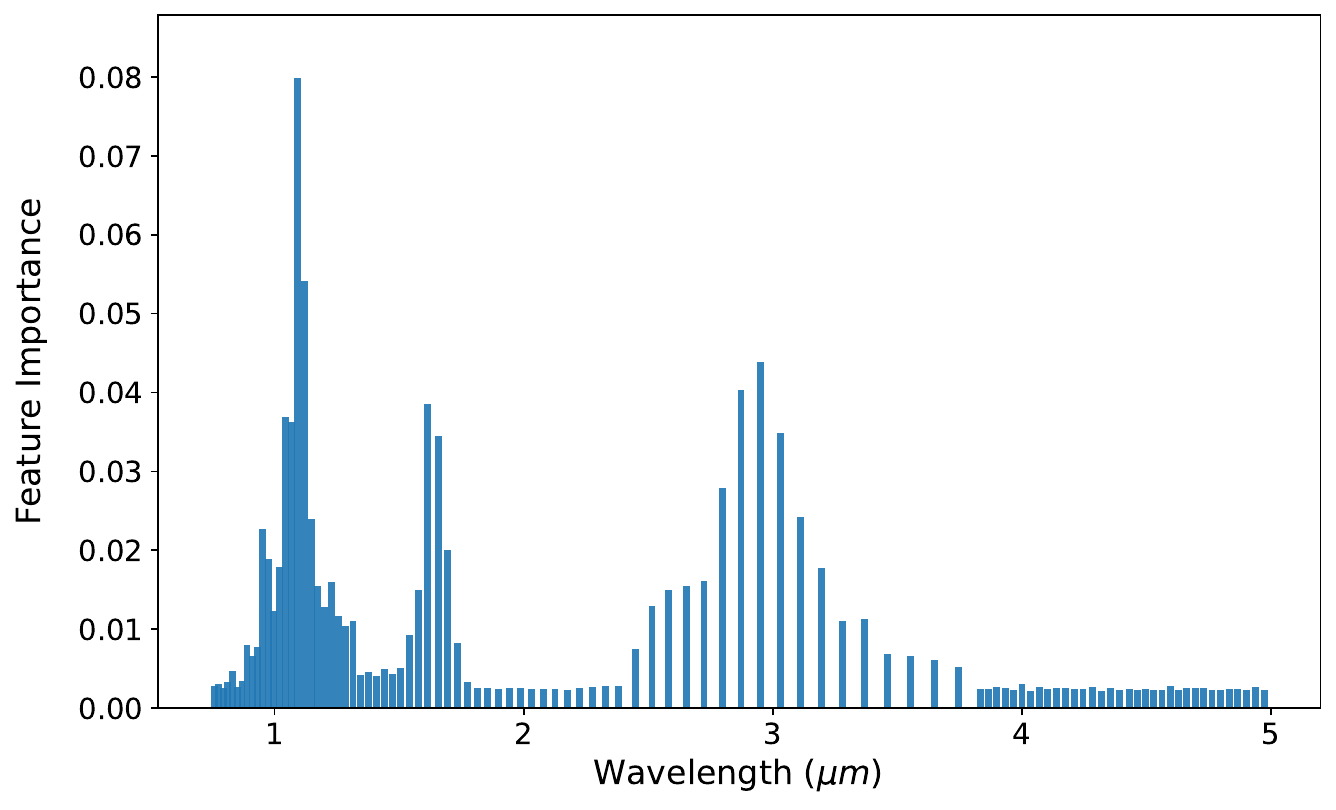}
    \caption{Feature importance as a function of observed wavelength, for a XGBoost classifier trained directly on spectra (\S \ref{sec:results}). The feature importance peaks at channels near 1.1, 1.6, and 2.9 $\mu$m, suggesting that the discriminative information is concentrated near a small number of spectral regions.}
    \label{fig:spectra_feature_importance}
\end{figure}
We perform stratified $k$-fold cross-validation on the validation set, setting $k=5$. For each partition, four subsets are used to train the XGBoost classifier, while the remaining held-out subset is used for testing. The held-out subset is changed across partitions, and the final metrics are averaged over the five test runs. Since XGBoost requires vector inputs, we flatten the unaligned spectrum encoder's $(5,128)$ intermediate features for each sample before training the classifier. For the unaligned image embeddings, we train an XGBoost classifier directly on the 1024-dimensional embeddings from AstroDINO. For the aligned image and spectrum encoders, we similarly train XGBoost classifiers directly on their respective 1024-dimensional embeddings. Finally, for the raw-spectra baseline, we train XGBoost on the 102 flux channels after normalizing each spectrum by its mean flux.

We summarize the performance of classifiers trained on each data representation in Table~\ref{tab:class_results}. For spectrum-based classification, alignment slightly improves performance relative to the unaligned spectrum encoder. However, the classifier trained on aligned spectrum embeddings still achieves lower stellar completeness and galaxy purity than the raw-spectra classifier, though the differences are within $2\%$. Improvements from alignment are considerably stronger for image-based classification, with the classifier trained on aligned image embeddings outperforming all other classifiers. The biggest gains are in stellar purity and galaxy completeness, which increase by  $\sim 17$ and $13$ percentage points, respectively, compared to the classifier trained on unaligned image embeddings. Stellar completeness and galaxy purity show more modest increases of $\sim 1.3$ and $1.0$ percentage points, respectively.

We show examples of star and galaxy image--spectrum pairs images alongside their XGBoost-derived galaxy class probabilities in Fig. \ref{fig:spec_img_proba}. The figure illustrates that imaging information can still yield substantially higher-confidence classifications when the spectral information is ambiguous or misleading. For example, the low-redshift galaxy in the bottom left panel has a low spectrum-based galaxy probability, but its extended structure in the imaging allows for a confidently correct classification. Conversely, in Fig. \ref{fig:high_spec_low_img}, we present examples of galaxy image-spectrum pairs that demonstrate how spectral information can help recover correct classifications in cases where imaging-based classification is incorrect. Although the sources appear compact and faint in imaging, redshifted spectral structure in the SPHEREx spectrophotometry provides sufficient discriminative information for confident galaxy classification.

We compare galaxy purity and completeness as a function of magnitude for classifiers trained on raw spectrum data and pre- and post-alignment spectrum encoder representations in Fig. \ref{fig:metric_comparison}. The classifiers are trained on data at the fiducial noise level and evaluated on test sets with both fiducial noise and a $\sqrt{2}\times$ higher RMS noise to assess robustness to differences in SNR. 

Classifiers trained on raw spectra and unaligned spectrum embeddings exhibit broadly similar performance across the full magnitude range. Under fiducial noise, galaxy purity remains above $99\%$ for both methods until $z_{\rm AB}\approx20$, before declining to $\sim96\%$--$97\%$ at the faint end. Completeness differences are larger, with the classifier trained on unaligned spectrum embeddings outperforming the raw-spectra classifier by $\sim9$ percentage points at bright magnitudes, though the results become comparable at the faint end. These bright bins contain a relatively small fraction of the galaxy sample (see Fig.~\ref{fig:data_distribution}), limiting their contribution to the aggregate completeness values in Table~\ref{tab:class_results}. The classifier trained on aligned spectrum embeddings further improves completeness marginally across mostly intermediate magnitudes ($z_{AB}\approx 20-21$), reaching gains of $\sim1$ -- $3$ percentage points relative to the unaligned-spectrum-embeddings classifier. At the same time, galaxy purity remains comparable to the raw-spectra baseline, with differences generally below $1.5$ percentage points. 

These trends persist under the higher-noise configuration. The raw-spectra classifier maintains a purity within $\sim2$ percentage points of the unaligned-spectrum-embeddings classifier, while the latter continues to achieve higher completeness by as much as $\sim10$ percentage points. The  classifier trained on aligned spectrum embedding retains approximately the same completeness gains over the unaligned-spectrum-embeddings and raw-spectra classifiers as in the fiducial-noise test. The three classifiers achieve nearly identical purity, differing from each other by less than $\sim2$ percentage points. Overall, the aligned and unaligned spectrum-based classifiers degrade by similar amounts under increased noise, indicating that alignment improves classification performance without reducing robustness to lower-SNR data.

To gain some insight into the most discriminative features in our spectra, we examine the feature importance of the raw-spectra classifier as a function of observed wavelength (Fig.~\ref{fig:spectra_feature_importance}). We use the default \texttt{gain} metric from XGBoost \citep{xgboost}, which computes the feature importance of a given input as the average reduction in the training objective achieved by splits using that feature, computed over all nodes and trees in the ensemble. As such, features with high importance are those that most effectively improve class separation. We find that the resulting importance distribution is highly concentrated, with bands centered on 1.1, 1.6, and 2.9~$\mu$m contributing most strongly to the classification, falling in SPHEREx bands 2$-$4. We recover the same wavelength peaks assuming several definitions of XGBoost feature importance \texttt{(weight, cover, total-gain, total-cover)} \footnote{Definitions for additional XGBoost feature importance metrics: \url{https://xgboost.readthedocs.io/en/latest/python/python_api.html}}. We likewise recover similar peaks by training a random forest classifier on the same dataset, for which feature importance is defined by the mean decrease in impurity (MDI, roughly analogous to the XGBoost \texttt{gain}). These tests establish the robustness of our derived feature importance wavelength peaks.

We interpret these observed wavelengths as corresponding to regions of the spectrum that vary strongly as function of redshift. The second peak in feature importance is aligned with the well-studied 1.6 $\mu$m bump, which corresponds to an opacity minimum of H$^-$ in stellar atmospheres \citep{16_bump}. This feature redshifts to higher observed wavelength for galaxies and remains fixed for Milky Way stars. The 1.1 $\mu$m and 2.9 $\mu$m peaks have less obvious associated spectral features but sample regions with important continuum slope information.  



\begin{figure*}
    \centering
    \includegraphics[width=\textwidth]{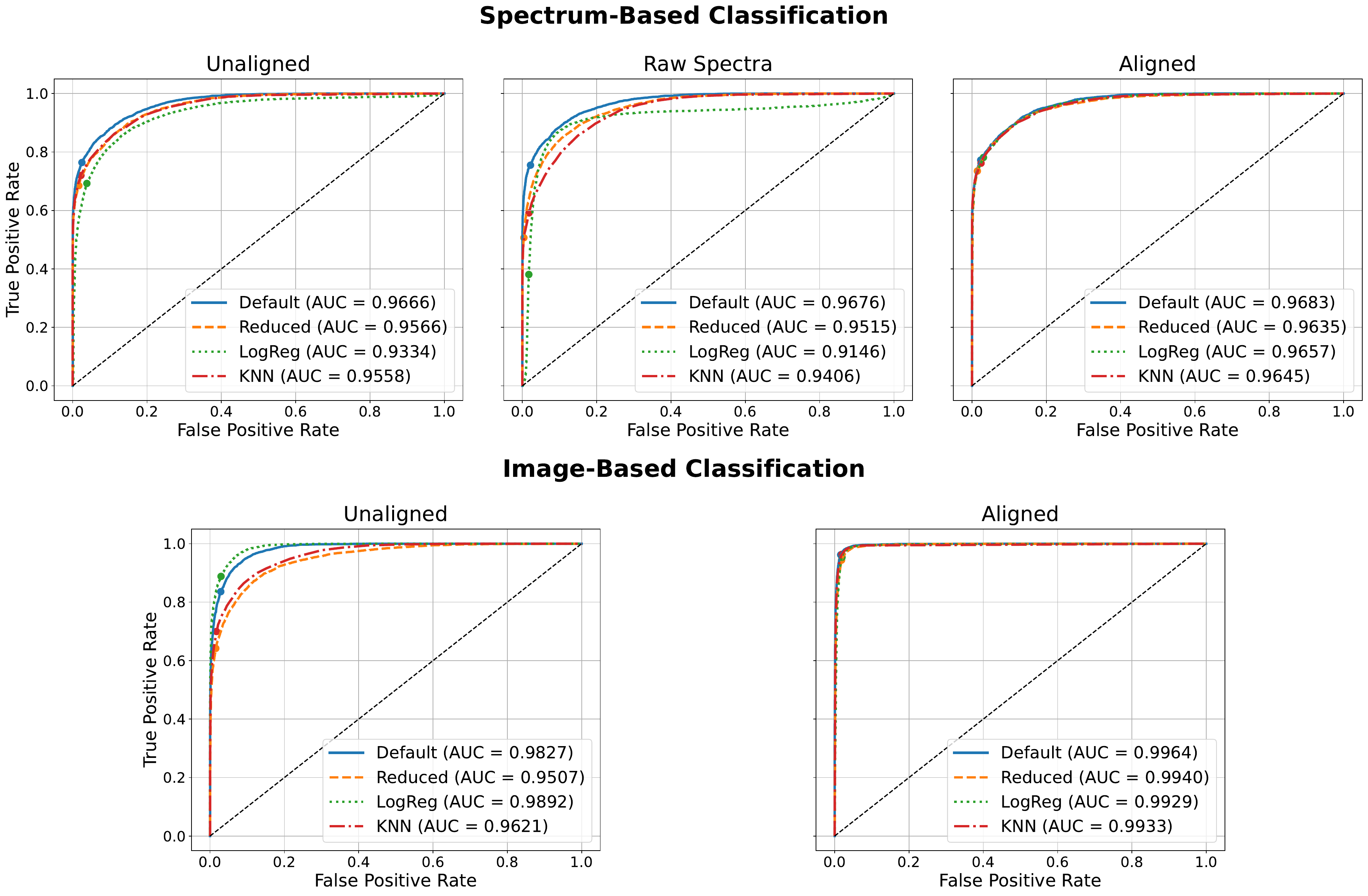}
    \caption{ROC curves for the XGBoost classifier. \textbf{Left:} embeddings from the unaligned spectrum encoder. \textbf{Middle:} classification performed directly on spectra. \textbf{Right:} embeddings from the aligned spectrum encoder. Curves are shown for our default XGBoost hyperparameters (max\_depth=6, n\_estimators=200), a reduced XGBoost model (max\_depth=3, n\_estimators=50), a logistic regression classifier, and a K-nearest neighbors classifier. Dots indicate the true positive rate (TPR) and false positive rate (FPR) at a classification threshold of 0.5.}
    \label{fig:roc}
\end{figure*}

\subsubsection{Classifier ablation tests}
\label{sec:alt_class}
Contrastive learning reorganizes the shared embedding space geometry so that representations of the same underlying physical object are brought into correspondence across modalities. Importantly, this process does not transfer information between modalities; rather, it reshapes how existing information is organized. In the limit of a sufficiently powerful classifier like our default XGBoost model, this restructuring may yield only marginal gains, as the classifier can already learn complex decision boundaries in the original space. In contrast, for simpler classifiers, the reorganized embedding spaces of the aligned encoders may offer significant advantages by revealing simpler decision boundaries. To test the significance of this effect, we evaluate separation performance using simpler classifiers trained on our different data representations. 

We compare our default XGBoost classifier against three simpler alternatives: (1) an XGBoost model with fewer trees ($200 \rightarrow 50$) and a shallower maximum tree depth ($6 \rightarrow 3$), (2) logistic regression with input features standardized to zero mean and unit variance, and (3) a k-nearest neighbors (KNN) classifier. Together, these models probe different aspects of representation quality: the reduced XGBoost model tests whether a less complex nonlinear decision boundary is sufficient, logistic regression evaluates whether the classes can be separated well by a simple linear boundary, and KNN measures how well class information is preserved in local neighborhood structure.

We compares their performances by ROC curves and AUCs in Fig. \ref{fig:roc}. For both the unaligned spectrum embeddings and the raw spectra, performance degrades noticeably when the weaker XGBoost and logistic regression classifiers are used, with the effect most pronounced for the raw spectra. Relative to the default XGBoost classifier, the reduced XGBoost, logistic regression, and KNN classifiers trained on the unaligned embeddings yield performance losses of $\Delta\mathrm{AUC}_{\rm Reduced}=-0.0100$, $\Delta\mathrm{AUC}_{\rm LogReg}=-0.0313$, and $\Delta\mathrm{AUC}_{\rm KNN}=-0.0108$, respectively. For the raw spectra, the corresponding values are $\Delta\mathrm{AUC}_{\rm Reduced}=-0.0161$, $\Delta\mathrm{AUC}_{\rm LogReg}=-0.0560$, and $\Delta\mathrm{AUC}_{\rm KNN}=-0.0270$, compared to the default raw-spectra XGBoost. In both cases, the KNN classifier outperforms the logistic regression, although both remain substantially below the default model. By comparison, the aligned spectrum embeddings exhibit much less performance loss under the same reduction in model complexity, with $\Delta\mathrm{AUC}_{\rm Reduced}=-0.0047$, $\Delta\mathrm{AUC}_{\rm LogReg}=-0.0028$, and $\Delta\mathrm{AUC}_{\rm KNN}=-0.0034$. Notably, the logistic regression classifier performs nearly as well as the default XGBoost model. In contrast, logistic regression produces the largest performance drop for both the unaligned embeddings and raw spectra. This suggests that the aligned embedding space is substantially more linearly separable and therefore less dependent on complex decision boundaries for effective classification.

We find similar results for image-based classification -- when trained on aligned embeddings, all alternative classifiers achieve performance within $\Delta\mathrm{AUC}<0.004$ of the default XGBoost model, while, in contrast, performance with unaligned image embeddings show stronger degradation ($\Delta\mathrm{AUC}_{\rm Reduced}=-0.032$ and $\Delta\mathrm{AUC}_{\rm KNN}=-0.0206$). Interestingly, logistic regression performs slightly better than the default XGBoost classifier in the unaligned case, achieving an AUC higher by 0.0065. This indicates that the latent space learned by the unaligned image encoder already provides a high degree of linear separability between stars and galaxies, with contrastive alignment further enhancing this structure.

Overall, these results are consistent with our expectations: alignment improves classification performance, with substantially larger gains for simpler classifiers. The reduced sensitivity of the aligned representations on classifier choice across both modalities indicates that class-discriminative information is more readily accessible after alignment.

\subsubsection{Interpreting improvements in image-based classification}
\label{sec:more_img}
\begin{figure*}
    \centering
    \includegraphics[width=0.48\textwidth]{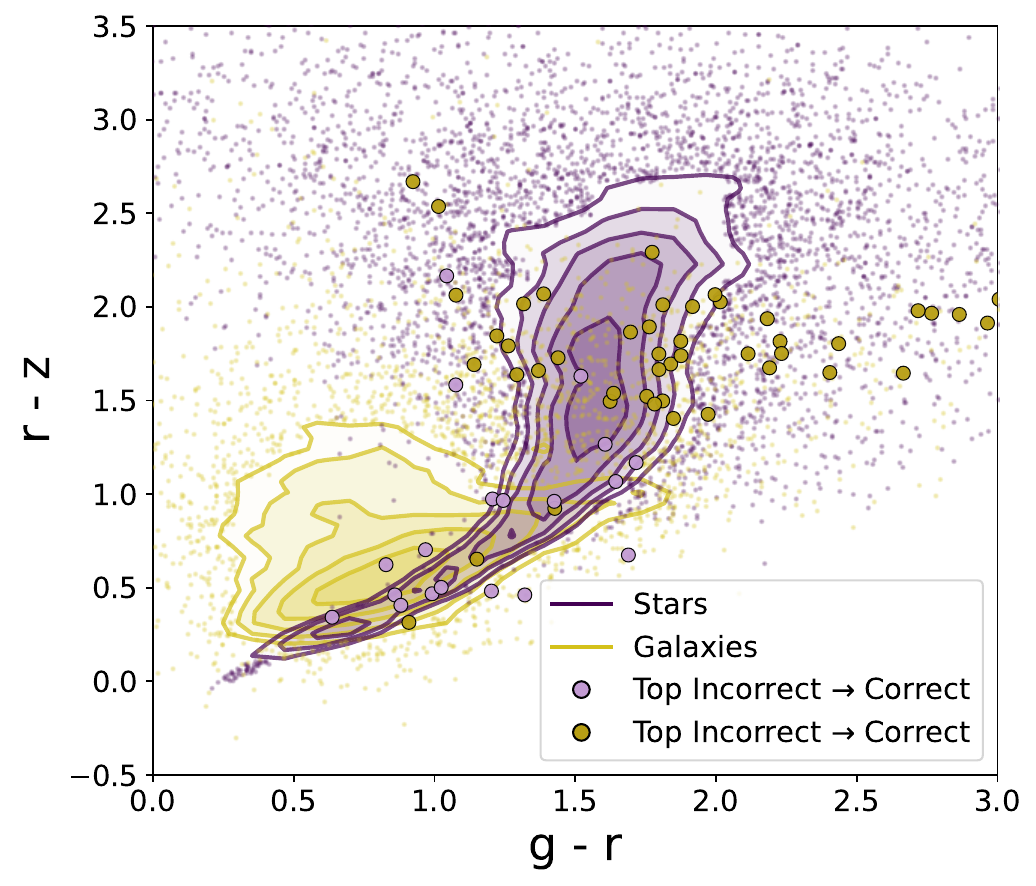}
    \includegraphics[width=0.48\textwidth]{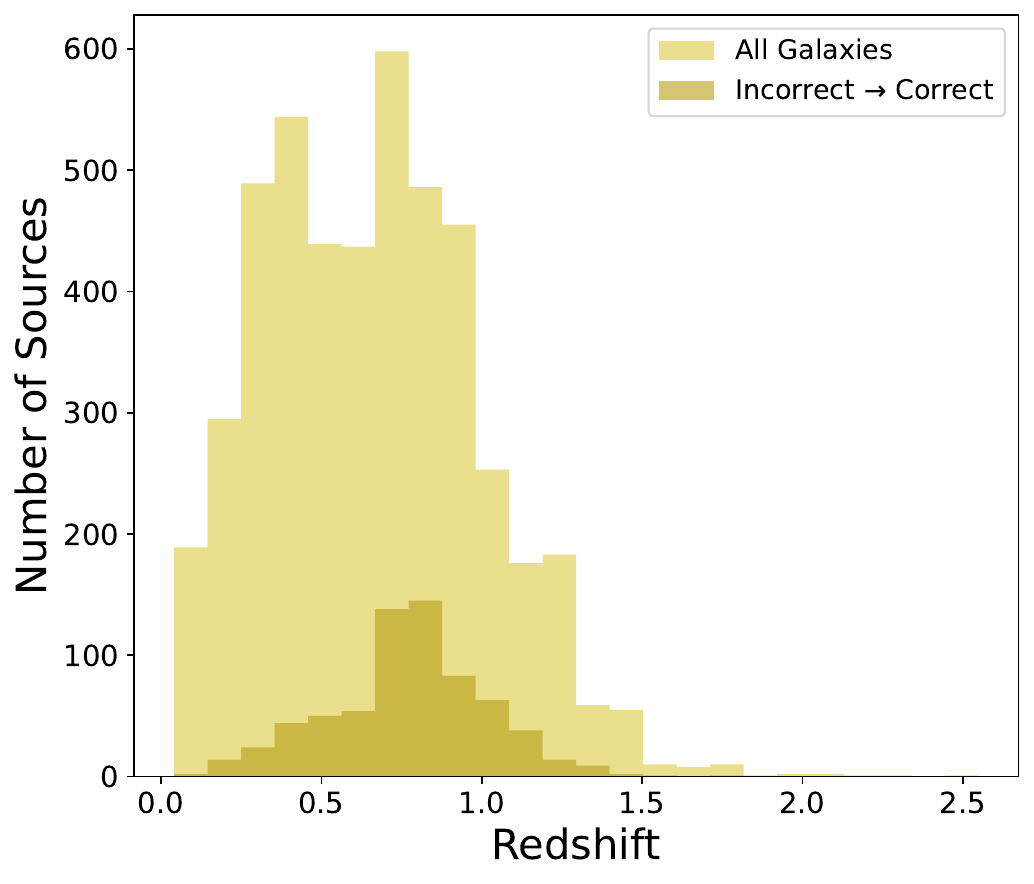}
    \caption{Properties of sources whose image-based classifications improve after contrastive alignment. \textbf{Left:} Optical color-color distribution of the full validation set, with the 70 sources showing the largest increase in classification probability for their true class after alignment highlighted. \textbf{Right:} Redshift distribution of all validation-set galaxies compared to galaxies that are misclassified before alignment but correctly classified afterward. The corrected sources are preferentially found at specific regions of color-color and redshift space, indicating that alignment mainly improves classification for particular subpopulations.
}
    \label{fig:most_improved}
\end{figure*}

For image-based classification specifically, we find that alignment improves performance dramatically, with nearly perfect classification results for our magnitude-limited sample (AUC$>0.992$ across all classifier variations). To investigate the cause of this improvement, we evaluate how alignment with the spectrum encoder alters the organization of the image embedding space.

\begin{table}[t]
\centering
\renewcommand{\arraystretch}{1.2}
\setlength{\tabcolsep}{12pt}
\caption{$R^2$ for linear regression models trained to predict observable properties from image embeddings. Values in parentheses correspond to the subset of sources that are incorrectly classified using unaligned image embeddings, but are correctly classified using aligned image embeddings.}
\label{tab:property_regression}
\fontsize{10.3}{12.8}\selectfont
\begin{tabular*}{\columnwidth}{@{\extracolsep{\fill}} lcc}
\hline
Property & Unaligned & Aligned \\
\hline
Redshift & 0.6253 ($-0.2804$) & \textbf{0.7782 (0.6833)} \\
$F(1.10\,\mu$m) & 0.2491 ($-0.1860$) & \textbf{0.3288 (0.3328)} \\
$F(1.60\,\mu$m) & 0.1905 ($-0.2148$) & \textbf{0.2555 (0.2213)} \\
$F(2.95\,\mu$m) & 0.3248 ($-0.2429$) & \textbf{0.4372 (0.4396)} \\
\hline
\end{tabular*}
\end{table}

We compute the coefficient of determination, $R^2$, for linear regression models trained to predict various observable properties from the unaligned and aligned image embeddings, with results summarized in Table \ref{tab:property_regression}. Numbers not in parentheses describe results for the full evaluation set, whereas numbers in parentheses correspond to the subset of sources that are incorrectly classified with their unaligned image embedding and correctly classified with their aligned image embedding.

Redshift is a natural quantity to examine, since all galaxies possess nonzero redshift while stars are fixed at $z=0$, meaning that stronger organization of the latent space according to redshift correlates with improved star--galaxy separability. Among the quantities tested, redshift exhibits the largest improvement in prediction accuracy after alignment, with $R^2$ increasing from 0.63 to 0.78. \citet{astroclip} report similar improvements in redshift regression using image embeddings after their own contrastive alignment procedure with DESI spectra. 

We do the same test for fluxes at the wavelengths corresponding to the strongest feature importance peaks identified by the classifier trained directly on raw spectra (Fig. \ref{fig:spectra_feature_importance}). Like in the classification tests, each spectrum's fluxes are normalized by its mean flux. Additionally, rather than taking the flux from the target wavelength, we average the flux across 5 channels, centered on the target wavelength. Here again, aligned embeddings consistently outperform unaligned embeddings, with the largest improvement occurring for the $2.95\,\mu\mathrm{m}$ flux. 

For the subsample of sources that have their classification corrected after alignment, we see a more dramatic improvement in $R^2$ between the unaligned and aligned image embeddings. The subsample $R^2$ values for the aligned representations remain comparable to the full sample values, whereas those for the unaligned representations they are essentially uncorrelated. Thus, for sources that are misclassified before alignment, several highly discriminative properties are substantially less accessible to a simple linear model in the unaligned representation. When these properties are well recovered from the aligned image embeddings, these sources are correctly classified. 

These results suggest a potential mechanism for the improvement in image-based classification. Contrastive alignment appears to reorganize information already present in the image embeddings such that redshift and spectral-shape information becomes more readily accessible. Because these properties are strongly discriminative between stars and galaxies, their improved organization within the aligned embedding space may contribute to the observed classification gains.

We further examine the sources whose image-based classification is corrected after alignment to determine whether the procedure uniformly improves classification or primarily helps certain subpopulations. In Figure \ref{fig:most_improved} we highlight the 70 sources with the largest increase in classification probability for their true class after alignment, shown in optical color-color space. We also show the redshift distribution of galaxies that are misclassified before alignment but correctly identified afterward, compared against the full galaxy sample. In color-color space, the corrected galaxies mostly cluster in a narrow region of $1.4 < r-z < 2.2$ with redder colors than the broader population and strong overlap with the main stellar distribution, indicating that alignment preferentially improves classification for a specific subset of galaxies. We see a similar trend for the improved star population, which is concentrated along the narrow blue stellar locus that heavily overlaps with the main galaxy distribution. The corrected galaxies primarily occupy redshifts $z \sim 0.7 - 0.9$, with the distribution skewing more towards higher redshift compared to the full sample. Galaxies at higher redshift are fainter and subtend smaller angular sizes, reducing the effectiveness of morphological information for classification at fixed angular resolution. The higher-redshift skew of the corrected sources is therefore consistent with the interpretation that alignment reorganizes the image embeddings to more strongly encode features correlated with spectral information, allowing these properties to play a larger role in classification.

To understand why spectrum-based classification exhibits more modest improvements pre- and post-alignment, we reemphasize that alignment can only restructure an embedding space according to properties that are correlated across modalities; information unique to one modality is not transferred to embeddings of the other. For spectrum-based methods, morphological information can be a strong complement, in particular for low-redshift galaxies that are most likely to be confused with stars. However, the spectrum embedding space has limited ability to reorganize around morphology during alignment because spectral information does not directly encode source shape. Therefore, the aligned spectrum encoder is unable to exploit a major discriminative property available within the imaging on its own.

\section{Implications for SPHEREx stellar contamination}
\label{sec:stellar_contam}

\subsection{Photo-z distributions of stars and galaxies}
\label{sec:photoz}
We now use the results from our star--galaxy separation tests to forecast stellar contamination for the SPHEREx All-Sky Galaxy Survey \citep{bock25}. For all sources, we run the template-fitting photo-z code developed in \cite{stickley16}, to determine which redshifts and tracer bins are most affected by star--galaxy classification errors. We deploy a model grid matching the resolution used in \cite{feder23}.

\begin{figure*}
    \centering
    \includegraphics[width=\linewidth]{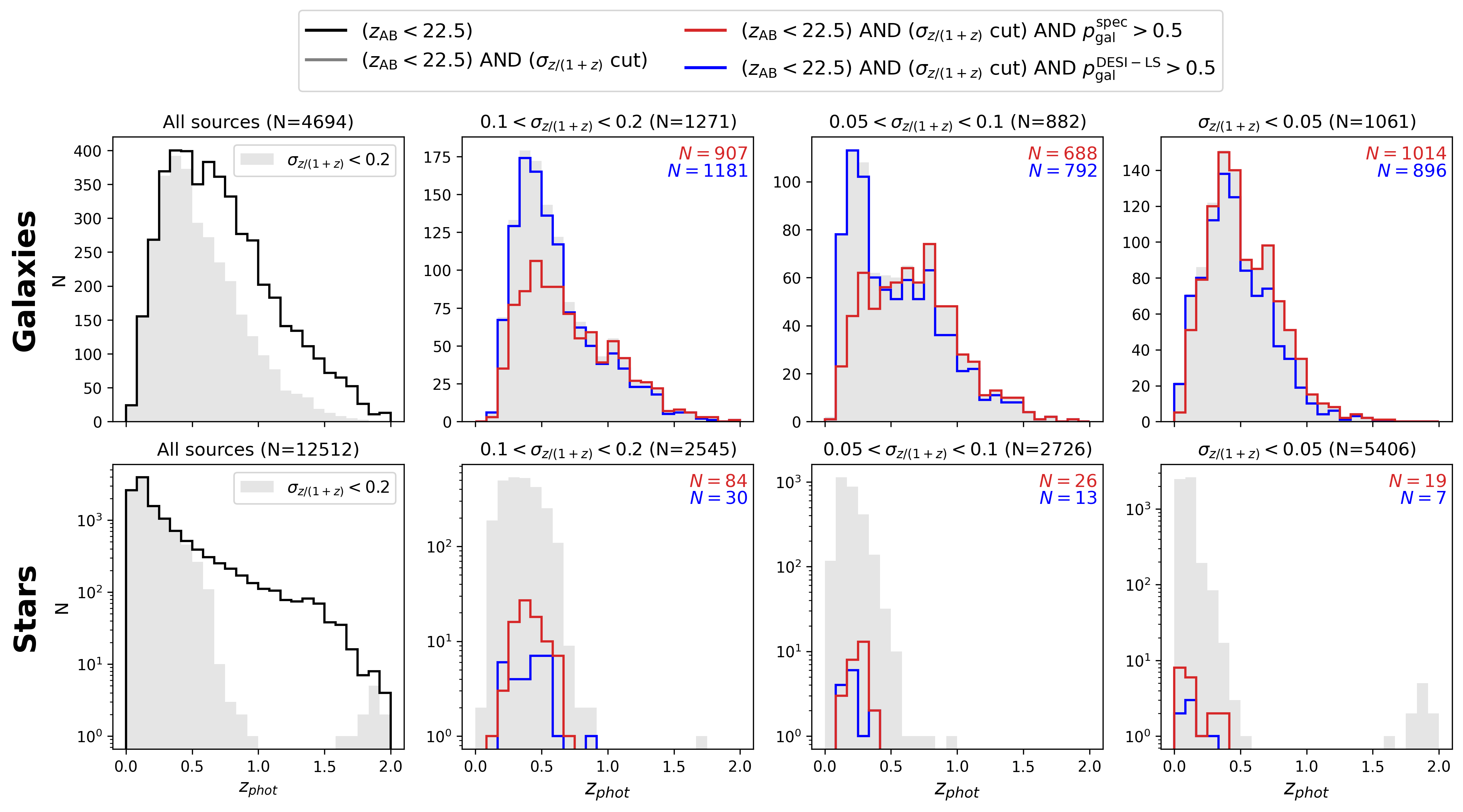}
    \caption{Impact of classification- and photo-z-based selections on simulated galaxies (top) and stars (bottom). In each panel we include the distribution of redshifts estimated using the template fitting code. Note that stars are plotted on a logarithmic y-axis scale.}
    \label{fig:photoz_dist_star_gal}
\end{figure*}

We first show the estimated redshift distributions of galaxies and stars in our synthetic dataset in Fig. \ref{fig:photoz_dist_star_gal}. In contrast to the galaxy redshift distribution, which is broad and extends to $z=2$, the estimated $p(z)$ distribution of the star sample strongly peaks at $z\sim 0$, with a tail also extending to $z=2$. While the $\sigma_{z/(1+z)}<0.2$ selection reduces the high-redshift tail of the galaxy sample, the impact of the cut on stars is significant: nearly all fictitious high-redshift solutions with $z > 1$ are removed. This follows the intuition that stellar SEDs resemble galaxies with redshifts $z \approx 0$. 

In the remaining panels, we plot galaxies and stars, binned by $\sigma_{z/(1+z)}$, with both spectrum- and image-based classifier selections applied (with $p_{\rm gal}^{\rm thresh} = 0.5$). Across the three bins, the two modalities are complementary for galaxy recovery. In particular, we find that the spectrum-based classifier is incomplete at low redshift ($z<0.6$) but retains nearly all high-redshift sources, while the opposite holds for the DESI-LS-based selection. This reinforces the picture that spectral degeneracies between stars and low-redshift galaxies drive the incompleteness at low redshift. We note that the DESI-LS contains both morphological information and some color information through broadband optical photometry (c.f. Fig. \ref{fig:data_distribution}). Both methods recover galaxies with low $\sigma_{z/(1+z)}$ (and correspondingly higher SNR) at higher rates, though performance differences between the two modalities persist. We likewise find that the number of retained stars decreases toward higher photo-z precision ($1-3\%$ of stars in the $0.1<\sigma_{z/1+z}<0.2$ bin and $<1\%$ in the higher-precision bins). The DESI-LS-based selection reduces the contaminant fraction by a factor of $\sim 2-2.5$ relative to the spectrum classification. 

\subsection{Galaxy completeness and purity}

\begin{figure*}
    \centering
    \includegraphics[width=0.85\linewidth]{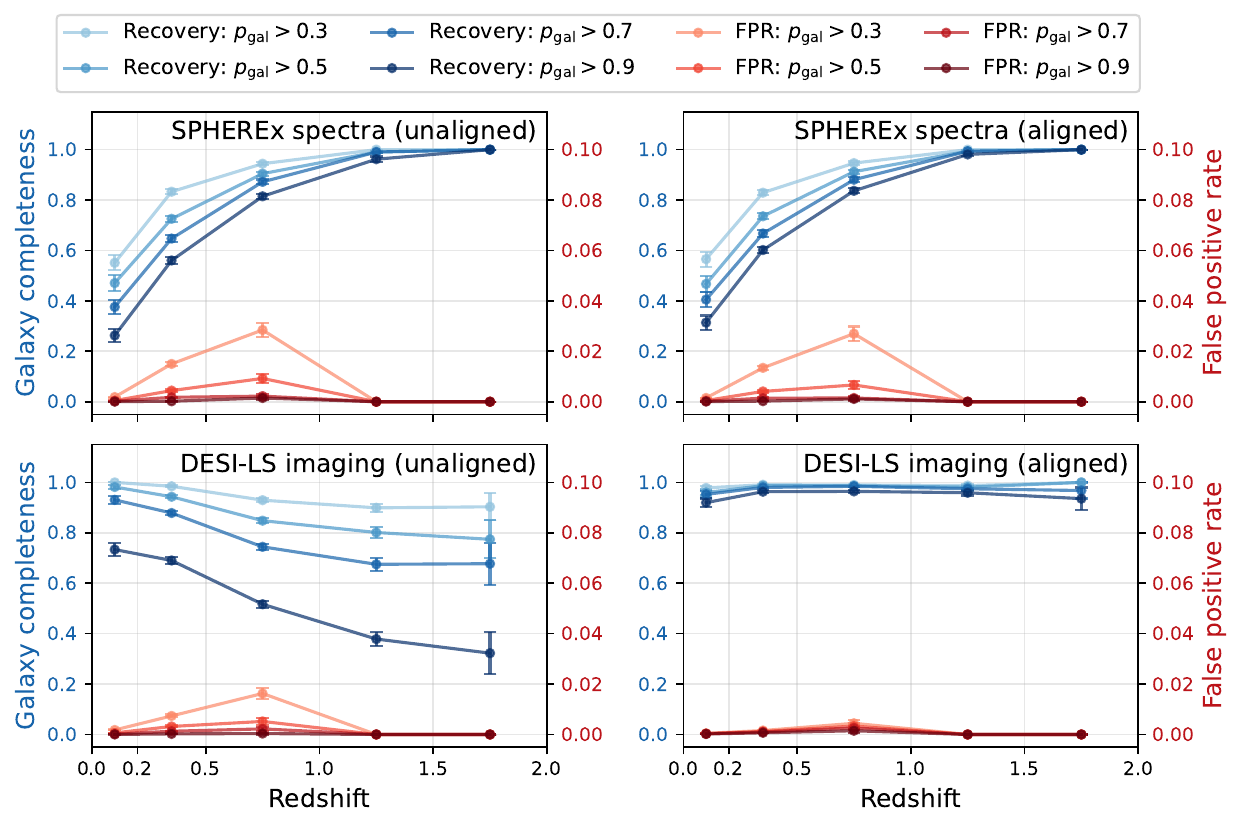}
    \caption{Impact of different classification probability thresholds for galaxy completeness (blue) and false positive rate (red) as a function of redshift, where here we define metrics for a sample with $z_{\rm AB}<22.5$ and $\sigma_{z/(1+z)}<0.2$ cuts applied. These false positive rate estimates assume stellar density of the COSMOS field.}
    \label{fig:comp_fdr_vs_redshift}
\end{figure*}

Rescaling the unnormalized results from \S \ref{sec:photoz} to stellar and galaxy densities, we compute the galaxy completeness and false positive rate (FPR) after restricting to a sample with $z_{\rm AB}<22.5$ and $\sigma_{z/(1+z)} < 0.2$. In Figure \ref{fig:comp_fdr_vs_redshift}, we show the resulting estimates as a function of redshift ranging from $0 < z_{\rm phot}<2$. For $z_{\rm AB}<22.5$, the stellar density in the COSMOS field is $\sim 12600$ deg$^{-2}$, while the galaxy density is $24500$ deg$^{-2}$ for the same selection. By computing these metrics with four different probability thresholds, $p_{\rm gal}^{\rm thresh} \in \lbrace 0.3, 0.5, 0.7, 0.9 \rbrace$, we see the expected tradeoff: completeness degrades with higher $p_{\rm gal}^{\rm thresh}$, while FPR improves (decreases). 

Notably, while the differences in the spectrum-based classifications are marginal between aligned and unaligned cases, we find that alignment significantly improves classification performance using the DESI-LS imaging, with nearly perfect source recovery and FPR across the full redshift range. 

\subsection{Forecasting contamination across the SPHEREx footprint}
\label{sec:forecast_contam_sky}
The presented results thus far assume galaxy and stellar densities from the COSMOS field, which resides at relatively high Galactic latitude. However, stellar density varies by a factor of $5-10$ over common extragalactic footprints. As such, we rescale our results using the variation in \emph{Gaia} stellar density (\texttt{NSIDE}=128) relative to the COSMOS field, in order to estimate the stellar contamination rate as a function of sky position.

\begin{figure*}
    \centering
\includegraphics[width=0.5\linewidth]{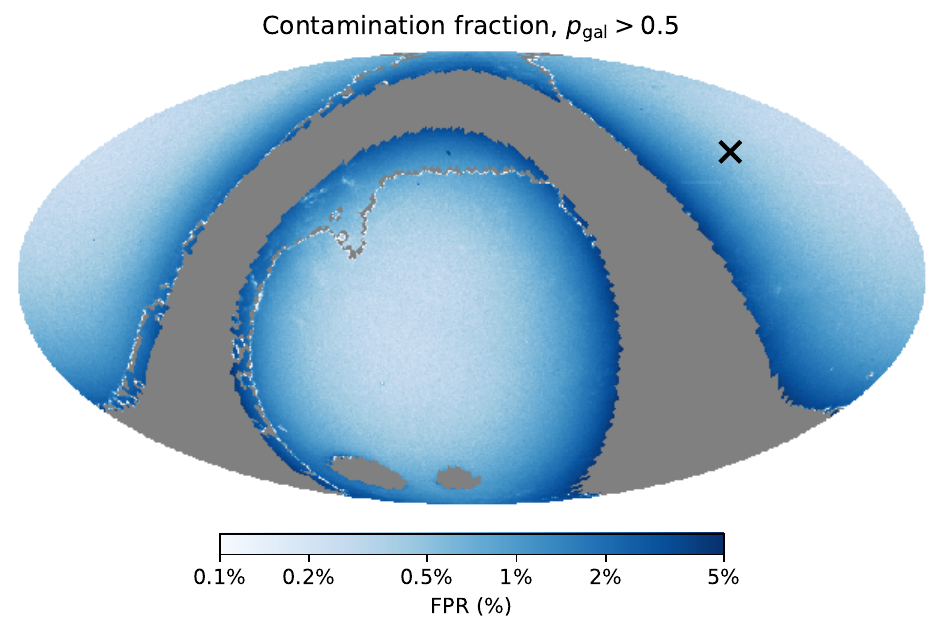} \includegraphics[width=0.49\linewidth]{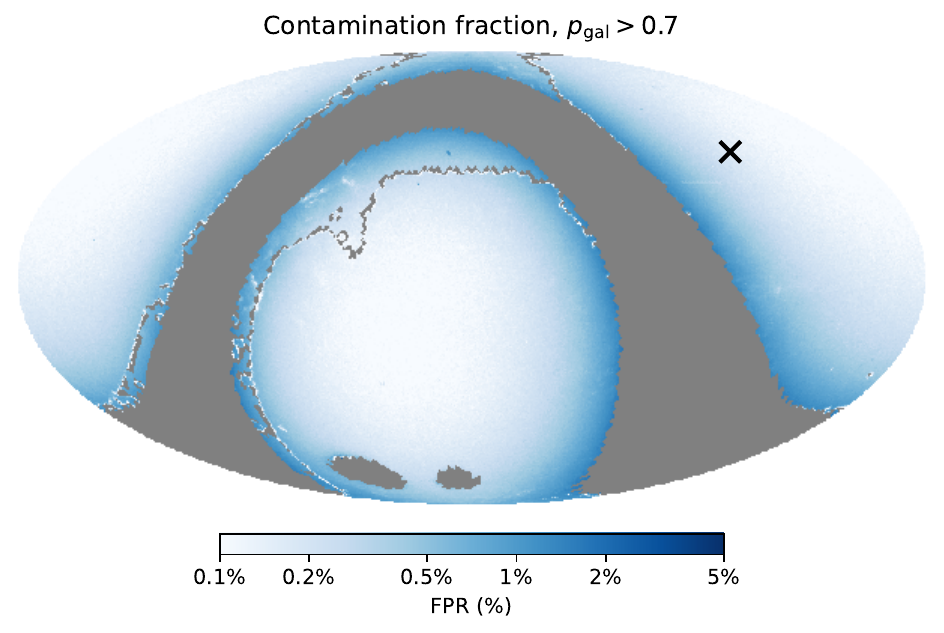}\includegraphics[width=0.49\linewidth]{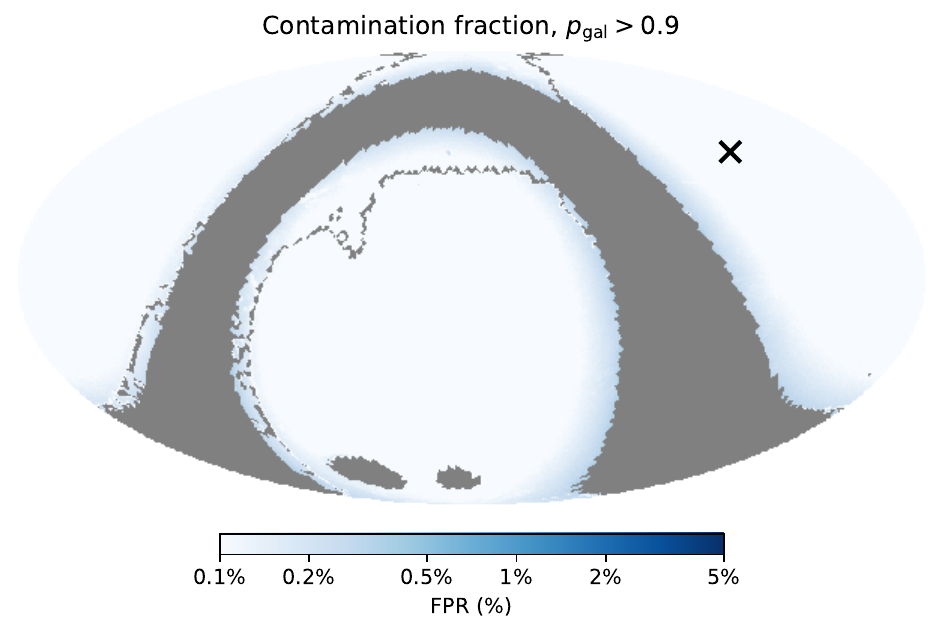}
    \caption{Forecasted stellar contamination fraction as a function of sky position, for different classification probability thresholds. In each case we include a notional mask that excludes regions in the Galactic plane, as well as additional masks for the LMC, SMC, etc. The black cross indicates the position of the COSMOS field. These results extrapolate classification performance from the aligned, spectrum-based embeddings in \S \ref{sec:results}.}
    \label{fig:contam_frac_map}
\end{figure*}

In Figure \ref{fig:contam_frac_map} we plot the contamination fraction $\eta_{\rm star}$ as a function of sky position, using the aligned spectrum-based results. The Galactic plane is masked, as well as other known extended objects (LMC, SMC), however there is a clear dependence of the contamination fraction on Galactic latitude. For our baseline selection, $p_{\rm gal} > 0.5$, $\eta_{\rm star}$ has a median of 1.3\%, with 90\% of the footprint having $\eta_{\rm star}<2.2\%$ for 90\% of HEALPix tiles and $<3.6\%$ for 99\% of tiles. However, as discussed in the previous subsection, there is a tradeoff between completeness and purity that can be leveraged through the exact choice of threshold $p_{\rm gal}$. For $p_{\rm gal} > 0.7$, $\eta_{\rm star}$ has a median of 0.4\%, with $\eta_{\rm gal}<0.9\%$ and 1.2\% for the same 90/99\% of tiles, respectively. For the same variation in $p_{\rm gal}$ from 0.5 to 0.7, the galaxy completeness degrades by only 5\% (for a $\sigma_{z/(1+z)}<0.2$ selected sample). These tradeoffs are reflected in Fig. \ref{fig:comp_fdr_vs_redshift}.

\subsection{Forecast caveats}

While the results in this section give some sense of the estimated systematic contamination from stars for the SPHEREx galaxy survey, our testing setup and associated forecasts use a number of simplifying assumptions:
\begin{itemize}
    \item \emph{Noise inhomogeneity}: Our simulated SPHEREx spectra assume a sky-averaged Zodiacal light photon noise level (as in \cite{feder23}), however the Zodiacal light intensity varies by a factor of $\sim 3$ as a function of ecliptic latitude. Regions near the Galactic plane will have higher overall backgrounds, also degrading the photon noise. As such, both the completeness and purity will vary as a function of sky position. By the same token, variations in the DESI-LS image seeing and residuals from bright star removal will impact the image-based information.
    \item \emph{Source blending}: While the DESI-LS images contain real galaxies with neighbors, the corresponding SPHEREx spectra are simulated without source confusion. In addition, our synthetic stars are injected at blank-sky positions that intentionally avoid catastrophic blends with other sources within our selection (i.e., $\theta_{\rm min} = 1.5\arcsec$). While such effects are expected to degrade classification performance, a dedicated study is beyond the scope of this work. 
    \item \emph{Fidelity of source injections}: we inject point sources into the DESI-LS maps using the provided PSF models, however any PSF errors may artificially impact the quality of our image-based classification. 
    \item \emph{Stellar density as traced by Gaia}: \emph{Gaia} detects and classifies sources using optical information; however, the types of stars that leak into galaxy samples may have a different spatial distribution than those of the stellar types that dominate the \emph{Gaia} sample.
\end{itemize}
\section{Conclusion}
\label{sec:conclusion}

In this work, we explore multimodal contrastive learning for star--galaxy separation by combining SPHEREx low-resolution spectrophotometry and DESI Legacy Survey imaging in a shared embedding space. We align image and spectrum encoders using a CLIP-style contrastive loss, and then use the resulting representations to train XGBoost classifiers for downstream star--galaxy classification. Alignment improves performance across both modalities, with the largest gains seen in the image-based representations. We apply these classifiers to a synthetic catalog matched to the COSMOS field and extrapolate the results across the full extragalactic footprint to assess the feasibility of achieving sub-percent stellar contamination in the SPHEREx All-Sky Galaxy Survey.

We additionally compare the separability of the different representations using classifiers of varying complexity. The aligned embeddings show substantially less degradation as classifier strength is reduced, suggesting that contrastive alignment restructures the embedding space in a way that makes discriminative information more accessible to simple classifiers. This simpler separation may improve robustness to observational systematics that perturb the embeddings. When the decision boundary is highly irregular, the same perturbation can affect different source populations differently, causing source completeness and purity to vary in a complicated manner as a function of survey variations. The linear separation enabled by alignment may therefore reduce the sensitivity of classification performance to such systematic effects.

After performing photo-z estimation on both stars and galaxies in our sample and applying additional redshift precision cuts, we find that stellar contamination can be controlled at the sub-percent level across most of the extragalactic sky. For a baseline selection of $p_{\rm gal}>0.5$, the median contamination fraction is 1.3\%; using a stricter $p_{\rm gal}>0.7$ cut reduces this to 0.4\%, at a cost of only 5\% in galaxy completeness. When using spectrum-based classification, the completeness loss is concentrated at low redshift, where star--galaxy degeneracies are most severe but also where the SPHEREx galaxy samples are most likely to remain sample-variance limited even for a reduced sample. Our testing setup can be used to inform selections that minimize stellar contamination errors and balance trade-offs between galaxy completeness and purity given more realistic degrees of variation.

Validating and extending these methods will require testing on real SPHEREx spectra and imaging. The DESI Legacy Survey forms a major component of the SPHEREx reference catalog \citep{akeson_spherex, yujin_refcat}, and objects with classifications from \emph{Gaia} and \emph{Euclid} will provide a direct means of validating star--galaxy separation across the magnitude range relevant to SPHEREx. The SPHEREx data itself provides an invaluable resource for refining stellar models \citep{zafar_spherex, ultracool_dwarfs} that can be compared against existing models such as TRILEGAL \citep{trilegal} and Galaxia \citep{galaxia}, with sufficient number statistics to reveal trends between dominant stellar type and sky location.

The multimodal framework presented here, along with the growing infrastructure of the MMU project provide a foundation for extending these methods to other ground-based imaging datasets such as Hyper-Suprime-Cam imaging and the \emph{Rubin} LSST, along with potentially other data modalities. More broadly, contrastive alignment may be of value for astronomical classification tasks beyond star--galaxy separation: in settings where relevant features are distributed across complex, nonlinear structures in high-dimensional data, alignment can reorganize these representations so that task-relevant variations are more explicitly encoded, reducing the burden on downstream models. As foundation models such as AION \citep{aion} continue to integrate heterogeneous astronomical data types, understanding the tradeoffs between cross-modal alignment and modality-specific information preservation will be an important consideration for optimizing such models for downstream tasks.

\section*{Acknowledgements}

R.M.F. is supported by NASA TCAN (grant number 80NSSC24K0101) and American Science Cloud grant ``DE-SCL0000092: HEP AmSC IDA Pilot: AI Universe". We thank Yun-Ting Cheng, Jamie Cheshire, Olivier Doré,  Zhaoyu Huai, Liam Parker, Natali di Santi, Joshua Speagle, and members of the SPHEREx science team for helpful discussions and feedback on the work in this manuscript. 

The DESI Legacy Imaging Surveys consist of three individual and complementary projects: the Dark Energy Camera Legacy Survey (DECaLS), the Beijing-Arizona Sky Survey (BASS), and the Mayall z-band Legacy Survey (MzLS). DECaLS, BASS and MzLS together include data obtained, respectively, at the Blanco telescope, Cerro Tololo Inter-American Observatory, NSF NOIRLab; the Bok telescope, Steward Observatory, University of Arizona; and the Mayall telescope, Kitt Peak National Observatory, NOIRLab. NOIRLab is operated
by the Association of Universities for Research in Astronomy (AURA) under a cooperative agreement with the National Science Foundation. Pipeline processing and analyses of the data were supported by NOIRLab and the Lawrence Berkeley National Laboratory. Legacy Surveys also uses data products from the Near-Earth Object Wide-field Infrared Survey Explorer (NEOWISE), a project of the Jet Propulsion Laboratory/California Institute of Technology, funded by the National Aeronautics and Space Administration. Legacy Surveys was supported by: the Director, Office of Science, Office of High Energy Physics of the U.S. Department of Energy; the National Energy Research Scientific Computing Center, a DOE Office of Science User Facility; the U.S. National Science Foundation, Division of Astronomical Sciences; the National Astronomical Observatories of China, the Chinese Academy of Sciences and the Chinese National Natural Science Foundation. LBNL is managed by the Regents of the University of California under contract to the U.S. Department of Energy.

\section*{Software}
This work has made use of the following packages: \texttt{matplotlib} \citep{matplotlib}, \texttt{numpy} \citep{numpy}, \texttt{scipy} \citep{scipy}, \texttt{PyTorch} \citep{pytorch}, \texttt{umap-learn} \citep{umap}, \texttt{astropy} \citep{astropy}, and \texttt{Hugging Face}.

\newpage
\appendix
\section{Image-Spectrum Alignment Visualization}
\label{appendix:alignment visualization}

\begin{figure}[hb]
    \centering
    \includegraphics[width=0.9\linewidth]{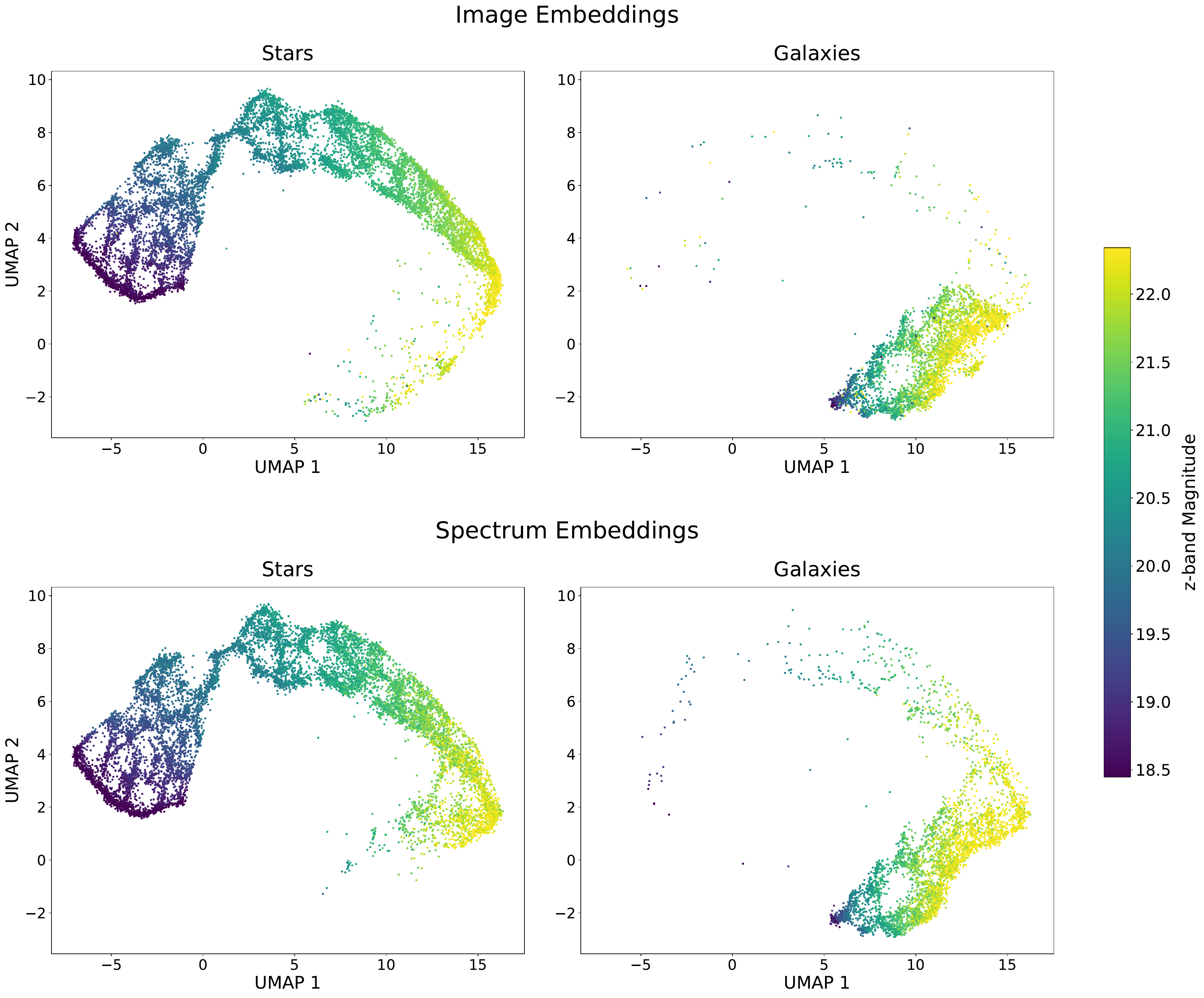}
    \caption{UMAP visualization of encoder embedding spaces. \textbf{Top:} aligned image embeddings of stars and galaxies. \textbf{Bottom:} aligned spectrum embeddings of stars and galaxies. While the contrastive objective encourages spectra and imaging of the same sources to reside at the same position in the shared embedding space, some discrepancies remain (see text).}
    \label{fig:umap_image_spectrum_zmag_comparison}
\end{figure}

In Figure \ref{fig:umap_image_spectrum_zmag_comparison} we compare the embeddings of the aligned image and spectrum encoders. While the manifold structure is broadly similar, differences emerge at both the faint and bright ends of the distributions. 

For galaxies, the image embedding space contains relatively few sources with UMAP-2 greater than 3 ($\sim2\%$ of all galaxies). In contrast, the spectrum embedding space shows a broader distribution of faint galaxies extending into regions dominated by faint stars. These galaxies comprises approximately $\sim 11\%$ of the galaxy sample. For stars, the image embedding manifold narrows to a sharp endpoint at faint magnitudes. In the corresponding region of the spectrum manifold, the distribution first broadens and is then largely truncated. For the spectrum embedding space, approximately $10\%$ of the star sample lies within the region where the stellar manifolds diverge ($5 <$ UMAP-1 $< 16$ and $-3.5 <$ UMAP-2 $< 2.5$), compared to $\sim6\%$ for the image embeddings space.

Smaller differences are also present for the brighter sample. For example, more bright galaxies appear shifted toward regions associated with bright stars in the spectrum embedding space compared to the image space. This difference suggests that alignment does not fully erase modality-specific structure: image and spectrum embeddings may still emphasize different physical properties of the same sources. The image embedding space is expected to be more sensitive to morphological information, making bright, extended galaxies easier to distinguish from stars. In contrast, the spectrum embedding space may organize some bright, likely low-redshift galaxies more similarly to stars if their broad spectral characteristics are less distinctive in the SPHEREx spectrophotometry. This interpretation is broadly supported by Fig.~\ref{fig:field_umap_comparison}, where galaxies lying closer to the stellar manifold tend to occupy lower redshifts.

Together, these differences suggest that the two embedding spaces may retain distinct inductive biases despite alignment. Consequently, image- and spectrum-based approaches may be susceptible to different failure modes, with misclassifications arising from different source properties depending on the modality.

\section{Embedding Space Organization Along Physical Properties}
\label{appendix:physical_properties}

\begin{figure}[h]
    \centering
    \includegraphics[width=\textwidth]{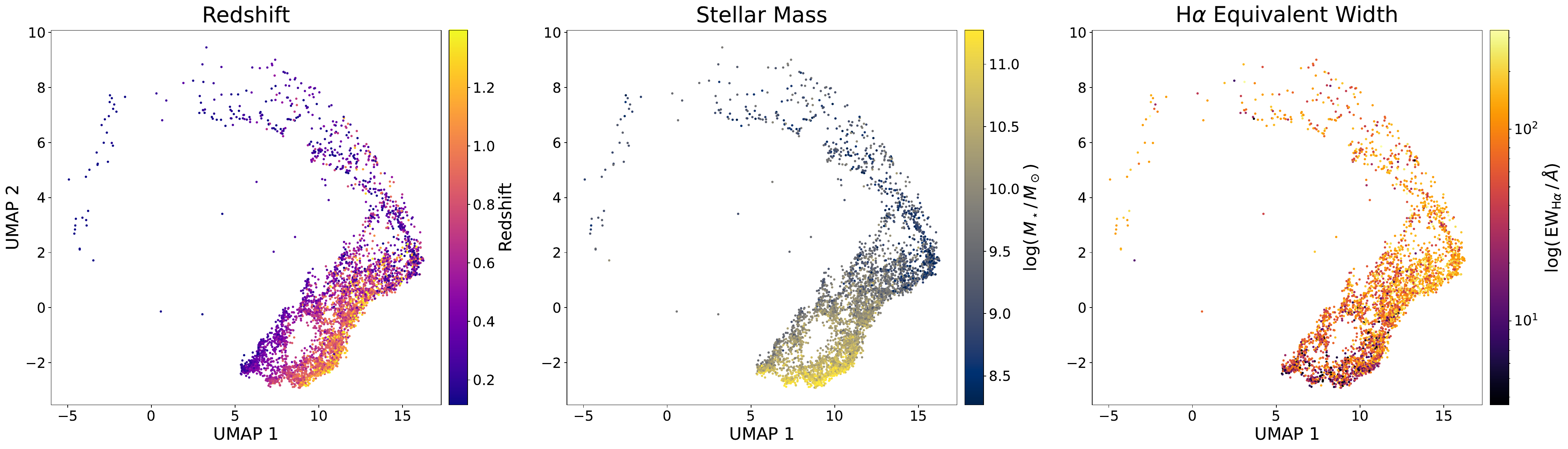}
    \caption{UMAP visualization of galaxy embeddings from the aligned spectrum encoder. \textbf{Left:} points colored by redshift. \textbf{Middle:} points colored by stellar mass. \textbf{Right:} points colored by H$\alpha$ equivalent width. The embedding exhibits clear organization with redshift and stellar mass, while the relationship with H$\alpha$ equivalent width appears noisier, though some structure remains visible.}
    \label{fig:field_umap_comparison}
\end{figure}

While alignment may improve the organization of the embedding space for star–galaxy classification, it is also informative to examine how galaxies alone are structured with respect to other physical properties. For the COSMOS galaxies in our dataset, we use redshift and stellar mass measurements from \cite{cosmos2020}, which are then used within the semi-empirical model of \cite{feder23} to predict H$\alpha$ equivalent width. Figure \ref{fig:field_umap_comparison} shows the aligned spectrum embeddings for the galaxies from Fig. \ref{fig:class_proba_umap_comparison}, colored by the three quantities above. The embedding space exhibits clear organization with respect to both redshift and stellar mass, indicating that galaxies with similar properties occupy similar regions of the latent space. The predicted H$\alpha$ equivalent width also shows similar structure, given its dependence on stellar mass and redshift \citep{feder23}.

We repeat the test done in \S \ref{sec:more_img} where we train a linear model to predict properties from the aligned spectrum embeddings. The regression results largely agree with the figure for all quantities shown. For redshift, the linear model achieves an $R^2 = 0.7618$. For stellar mass, it achieves a similar value of $R^2 = 0.7652$. As expected, for H$\alpha$ equivalent width, it achieves a lower value of $R^2 = 0.2415$ .


\bibliography{references}{}
\bibliographystyle{aasjournal}

\end{document}